\documentclass[10pt,twocolumn,letterpaper]{article}

 \usepackage{cvpr}              % To produce the CAMERA-READY version
\usepackage{algorithm}
\usepackage{algpseudocode}  % replaces algorithmic
\usepackage{dsfont}
\usepackage{booktabs}
\usepackage{multirow}
\usepackage[normalem]{ulem}
\newenvironment{red}{\color{red}}{}

\newenvironment{blue}{\color{blue}}{}
\newcommand\remove{\bgroup\markoverwith{\textcolor{orange}{\rule[.5ex]{2pt}{1pt}}}\ULon}

\usepackage{graphicx}
\usepackage{tikz}
\usetikzlibrary{calc}
\usepackage{xcolor}
\usepackage{siunitx} % for \num{...}
\usetikzlibrary{positioning}
\usepackage{colortbl}
\newcommand{\scorebadge}[1]{%
  \node[anchor=south east, fill=black!65, text=white, rounded corners=1pt,
        font=\scriptsize, inner xsep=3pt, inner ysep=1.5pt] at (im.south east) {\num{#1}};
}
\newcommand{\dislikedbadge}{%
  \node[anchor=north west, fill=red!70, text=white, rounded corners=1pt,
        font=\scriptsize, inner xsep=3pt, inner ysep=1.5pt] at (im.north west) {Disliked};
}
\newcommand{\imgscore}[3][]{%
  \begin{tikzpicture}[baseline,outer sep=0]
    \node[inner sep=0] (im) {\includegraphics[width=0.19\linewidth,#1]{#2}};
    \scorebadge{#3}
  \end{tikzpicture}%
}
\newcommand{\imgscoredisliked}[3][]{%
  \begin{tikzpicture}[baseline,outer sep=0]
    \node[inner sep=0] (im) {\includegraphics[width=0.19\linewidth,#1]{#2}};
    \dislikedbadge
    \scorebadge{#3}
  \end{tikzpicture}%
}

\usepackage{amsmath}

\usetikzlibrary{calc,positioning}

\definecolor{accent}{HTML}{1D4ED8} % calm blue; grayscale-safe alternatives: 444444 or 374151

\newcommand{\ind}{\perp\!\!\!\!\perp} 

\definecolor{cvprblue}{rgb}{0.21,0.49,0.74}
\usepackage[pagebackref,breaklinks,colorlinks,allcolors=cvprblue]{hyperref}

\def\paperID{17800} % *** Enter the Paper ID here
\def\confName{CVPR}
\def\confYear{2026}

\title{Personalized Image Generation for Recommendations Beyond Catalogs}

\author{
Gabriel A. Patron \quad
Zhiwei Xu \quad
Ishan Kapnadak \quad
Felipe Maia Polo \\
University of Michigan \\
{\tt\small \{gapatron, xuzhiwei, kapnadak\}@umich.edu, felipemaiapolo@gmail.com}
}

\begin{document}
\maketitle

\begin{figure}[t]
  \centering
  \includegraphics[trim={4cm 5cm 4cm 4cm},clip, width=\linewidth, scale=0.75]{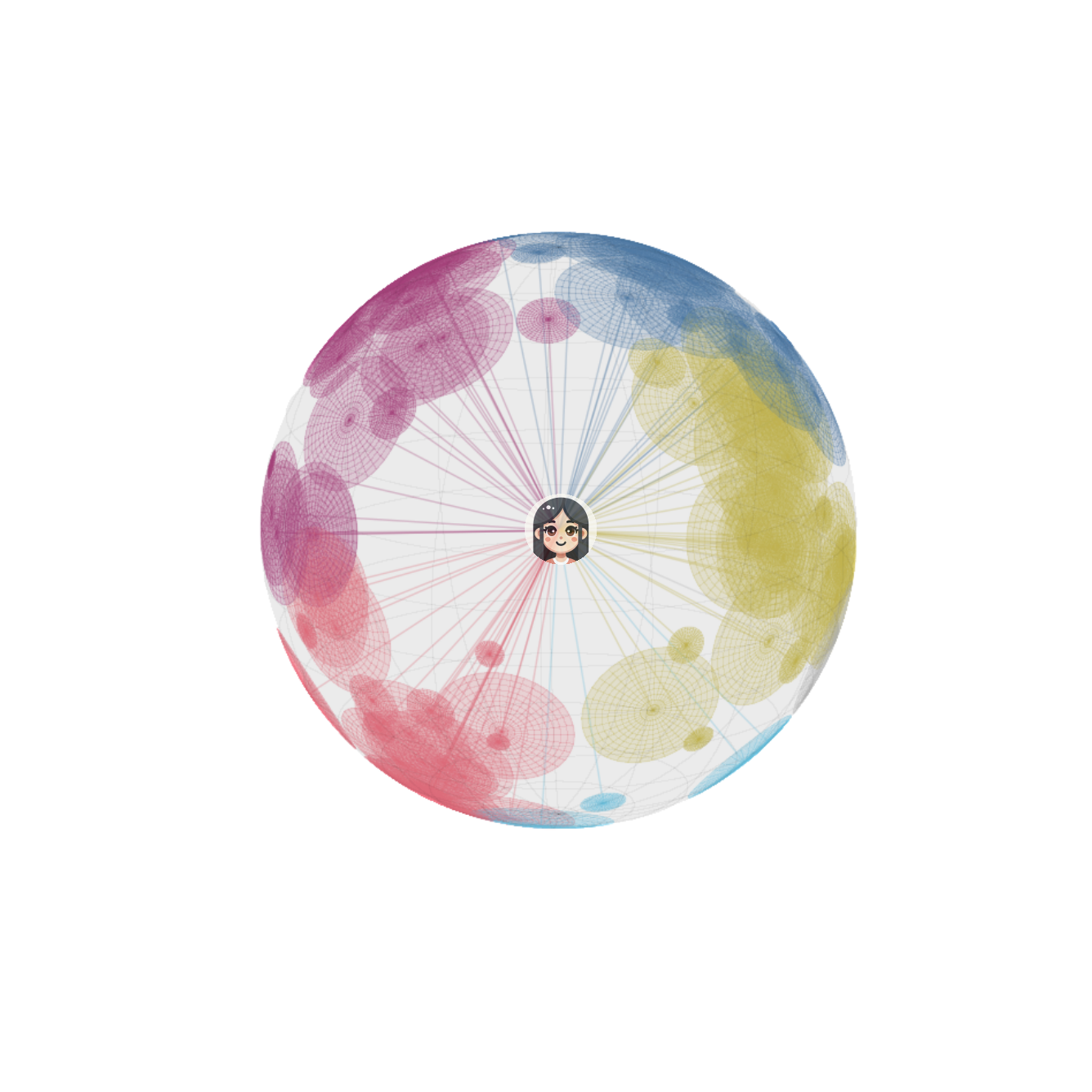}
  \caption{\texttt{REBECA} learns a user-conditioned diffusion prior whose geometry spans the preference manifolds of all users. Conditioning selects a region of this shared embedding space, from which the prior samples diverse yet personalized embeddings.}
  \label{fig:example}
\end{figure}
%%%
%%%
\textbf{Abstract.} Personalization is central to human-AI interaction, yet current diffusion-based image generation systems remain largely insensitive to user diversity. Existing attempts to address this often rely on costly paired preference data or introduce latency through Large Language Models. In this work, we introduce \texttt{REBECA} (REcommendations BEyond CAtalogs), a lightweight and scalable framework for personalized image generation that learns directly from implicit feedback signals such as likes, ratings, and clicks. Instead of fine-tuning the underlying diffusion model, \texttt{REBECA} employs a two-stage process: training a conditional diffusion model to sample user- and rating-specific image embeddings, which are subsequently decoded into images using a pretrained diffusion backbone. This approach enables efficient, fine-tuning-free personalization across large user bases. We rigorously evaluate \texttt{REBECA} on real-world datasets, proposing a novel statistical personalization verifier and a permutation-based hypothesis test to assess preference alignment. Our results demonstrate that \texttt{REBECA} consistently produces high-fidelity images tailored to individual tastes, outperforming baselines while maintaining computational efficiency\footnote{Code available in \href{https://anonymous.4open.science/r/REBECA-CCB6/README.md}{ anonymized repository during review.}}.

%We present \texttt{REBECA}, a scalable framework for personalized diffusion that learns user-conditioned image generation directly from implicit feedback such as likes or clicks. REBECA trains a conditional diffusion prior \ZX{maybe just say conditional difffusion model? We haven't defined `diffusion prior'} over CLIP embeddings $p_\theta(I^e\!\mid U,R)$ and decodes them with a frozen diffusion model, requiring no paired preference data, text prompts, or model retraining. Across 210 users, REBECA improves macro-F1 by $+7$ points and verifier-based personalization by $+5\sigma$ over LoRA and VLM baselines, while matching aesthetic quality. Training completes in $<0.2$ GPU-days on a single RTX 4090. These results demonstrate that implicit-feedback personalization can be both effective and computationally lightweight \footnote{We provide an \href{http://www.example.com}{anonymized \texttt{REBECA} repository}}. \FMP{I would try to make the abstract less technical (perhaps no need to mention CLIP and explicit baselines) and more engaging (and not use math notation)}
%\ZX{Yes numbers and technical details can be presented in experiments section or appendix}
    
\section{Introduction}

Personalization is central to human-AI interaction, as users exhibit diverse tastes, intents, and creative goals. However, current diffusion-based image generation systems remain largely insensitive to such user diversity, producing outputs that are visually impressive but not tailored. Addressing this limitation is crucial for advancing applications in creative tools, advertising, and content recommendation.

Recent work has begun to address personalization in diffusion models. \emph{Personalized Preference Fine-tuning (PPD)}~\cite{dang2025personalized} adapts diffusion models to user tastes through pairwise preference supervision, where users choose preferred images from pairs. While effective, this approach depends on costly paired data and model fine-tuning, limiting scalability. \emph{Personalized Multimodal Generation (PMG)}~\cite{shen2024pmg} instead leverages large language models (LLMs) to infer user preferences from behaviors such as clicks or chat histories, but introduces latency from LLM inference and relies heavily on textual cues rather than richer multimodal signals.

In this work, we introduce \texttt{REBECA} (REcommendations BEyond CAtalogs), a lightweight and scalable framework for personalized image generation with diffusion models. Unlike prior methods that rely on paired annotations or large language models, \texttt{REBECA} learns personalization directly from behavioral data such as likes, ratings, and clicks that are readily available in real-world social media platforms. The method operates in two stages:
\begin{enumerate}
    \item Train a conditional diffusion to model the distribution $p_{\hat{\theta}}(I_e\mid U,R)$. Then samples image embeddings $I_e$ conditioned on a user $U$ and rating $R$.
    \item Decode $I_e$ into personalized images $I$ using a pretrained diffusion model $p(I\mid I_e)$ that conditionally decodes the embedding.
\end{enumerate}
Unlike prior work, REBECA does not rely on textual descriptions, paired preference labels, or per-user fine-tuning. Instead, we learn a single conditional diffusion prior whose geometry spans the preference manifolds of all users. Conditioning identifies a user-specific region of this shared space, enabling scalable, plug-and-play personalization

We rigorously evaluate \texttt{REBECA} on real-world user datasets, measuring both personalization strength and visual fidelity. In addition to more traditional metrics such as recall, precision, and predicted image quality, we propose and employ a statistical personalization verifier and a permutation hypothesis test for personalized generations. Our results on synthetic and real data show that \texttt{REBECA} consistently produces images that align with individual user preferences. In summary, our contributions are threefold:
\begin{enumerate}
\item \textbf{Learning from implicit signals.} We introduce a single user-conditioned diffusion prior that models personalized CLIP-space embeddings directly from implicit feedback, eliminating the need for preference pairs, LLM mediation, or per-user fine-tuning.
\item \textbf{Lightweight, plug-and-play framework.} REBECA decouples personalization from the image generator: the learned prior produces embeddings that plug into any pretrained decoder with the same embedding class, enabling large-scale personalized generation with minimal compute.
\item \textbf{Rigorous evaluation.} We propose a statistical personalization verifier and a permutation-based hypothesis test to assess alignment with user preferences, complementing standard metrics, and show strong performance on both synthetic and real datasets. We believe our newly introduced evaluation procedure can be used by future work as a standard evaluation approach for personalized generations.
\end{enumerate}

\section{Related Work}
\begin{figure*}[t]
  \centering
  \vspace{-1em} % optional: pulls the figure up closer to the top margin
  \begin{minipage}[t]{\linewidth}
    \includegraphics[width=\linewidth,page=1]{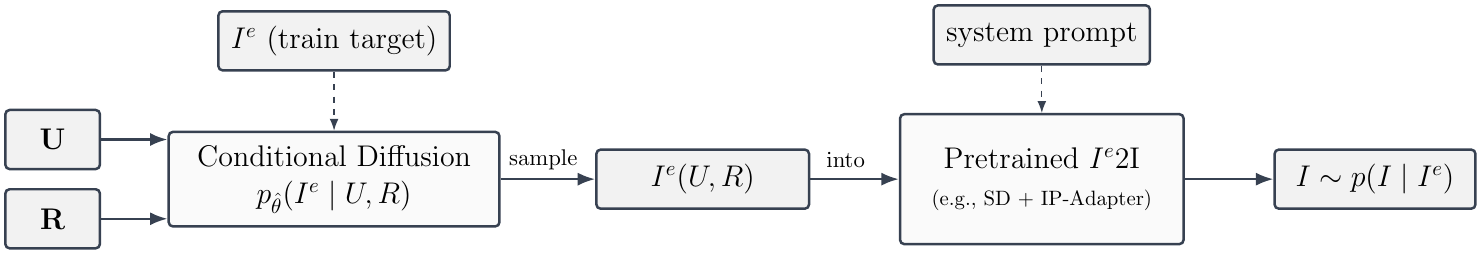}
  \end{minipage}
  \vspace{-0.5em} % optional: tighter space below
  \caption{
\textbf{\texttt{REBECA} overview.}
\textit{Training:} Conditional diffusion prior trained to generate personalized image embeddings from user IDs and ratings.  
\textit{Inference:} Generated embeddings are decoded into images via a pretrained image decoder model.
}
  \label{fig:architecture}
\end{figure*}

\vspace{0.1cm} \noindent \textbf{Parameter-Efficient Fine-Tuning (PeFT).}
Adapters have emerged as an efficient means of specializing large text-to-image
models without updating the full backbone~\cite{hu2021loralowrankadaptationlarge,
zhang2023adding,ye2023ip-adapter}. In diffusion models, IP-Adapter~\cite{ye2023ip-adapter}
extends controllability to visual references by injecting image-conditioned
cross-attention layers, while methods such as LoRA, Textual Inversion, and
DreamBooth~\cite{hu2021loralowrankadaptationlarge,gal2022image,ruiz2023dreambooth,mou-t2i}
provide instance-level personalization from a handful of example images.
However, these approaches require explicit reference content (captions or images)
and therefore do not apply directly to implicit-feedback or recommendation
settings where preferences must be inferred rather than provided as exemplars.

\texttt{REBECA} generalizes the adapter concept by conditioning the diffusion
prior on user embeddings learned from behavioral data, rather than textual or
visual exemplars. This yields a compact, reusable personalization layer that
captures user-level preference structure and scales across hundreds of users
without per-user fine-tuning.

\vspace{0.1cm}
\noindent\textbf{Diffusion Model Alignment via Reinforcement Learning.}
A complementary line of work aligns diffusion models to external objectives
using reinforcement learning. Several methods apply PPO-style optimization
to diffusion trajectories or denoising steps~\cite{black2024trainingdiffusionmodelsreinforcement,
fan2023dpokreinforcementlearningfinetuning,zhang2024largescalereinforcementlearningdiffusion},
using rewards from aesthetic models, human preference predictors, or task-specific heuristics.
Other approaches introduce reward-conditioned diffusion objectives~\cite{zhao2025addingconditionalcontroldiffusion}, 
or train reward models that guide sampling in a policy-improvement loop.

These methods operate on text prompts or global image-level rewards and typically
require dense, high-quality reward signals. In contrast, \texttt{REBECA} learns
from implicit feedback and models personalized embedding 
distributions directly, providing user-level alignment without policy iteration
or reinforcement learning.

\vspace{.1cm} \noindent \textbf{Personalized Preference Alignment.} More recent work introduces explicit preference alignment. 
\emph{Personalized Preference Fine-tuning of Diffusion Models (PPD)}~\cite{dang2025personalized} extends Direct Preference Optimization (DPO)~\cite{rafailov2024directpreferenceoptimizationlanguage} to the multi-user setting. 
PPD derives user embeddings from the hidden representations of a vision–language model (VLM) trained on a few pairwise preference examples per user. 
It then fine-tunes a decoupled cross-attention module under multiple reward objectives to align generation with those embeddings. 
While effective, PPD relies on paired preference data, repeated VLM evaluations per user, and additional sample generation for each forward pass, which severely limits scalability.

\emph{Personalized Multimodal Generation (PMG)}~\cite{shen2024pmg}, leverages large language models (LLMs) to infer user preferences from behavioral traces such as images, clicks, or conversations. PMG translates these multimodal behaviors into natural-language prompts, keywords, and embedding representations that condition a separate diffusion-based generator. However, this approach introduces latency and computational cost due to the LLM inference stage, and personalization remains largely mediated by text rather than by directly learning aesthetic alignment. 

\texttt{REBECA} differs fundamentally from these approaches: it dispenses with both preference-pair supervision and VLM-based mediation, learning personalization directly from implicit feedback signals such as likes, ratings, or clicks. By conditioning a diffusion prior on compact user embeddings, \texttt{REBECA} achieves scalable, plug-and-play personalization while preserving the generality of pretrained image generators.

\section{Methodology}

We present the \texttt{REBECA} pipeline, a two-stage framework for personalized image generation from implicit feedback. 
In training, images are first mapped to a lower-dimensional embedding space using CLIP~\cite{radford2021learningtransferablevisualmodels}. 
A conditional diffusion model~\cite{ho2020denoisingdiffusionprobabilisticmodels} is then trained on these embeddings using classifier-free guidance~\cite{ho2022classifierfreediffusionguidance}, conditioned on user and rating information.
At inference time, the trained embedding diffusion model is provided with a user ID and a desired rating to generate image embeddings, which are then decoded into images using a pre-trained image generator.  \texttt{REBECA} consists of two key components:
\begin{enumerate}
  \item \textbf{Personalized embedding generator.} 
  A conditional diffusion prior samples CLIP-space embeddings directly conditioned on user IDs and ratings.
  \item \textbf{Decoder model.}
  The generated embeddings are translated into images by a frozen text-to-image model (e.g., Stable Diffusion ~\cite{podell2023sdxlimprovinglatentdiffusion,rombach2021highresolution}) augmented with an IP-Adapter \cite{ye2023ip-adapter} to enable image-conditioned generation.
\end{enumerate}
See Figure \ref{fig:architecture} for an overview. These design choices address the limitations of prompt-based personalization and enable flexible, language-free conditioning in CLIP space.

\vspace{.1cm} \noindent \textbf{Personalized image generation.}
We aim to sample personalized images $I$ from $p(I \mid U,R)$, where $U$ denotes a user and $R$ a desired rating. 
We factorize:
\begin{align}
p(I \mid U,R)
&= p(I, I^e \mid U,R) \nonumber \\
&= p(I^e \mid U,R)\,p(I \mid I^e,U,R),
\label{eq:factor}
\end{align}
where $I^e$ is the deterministic CLIP embedding of $I$. We approximate the two conditional distributions as follow. First, a conditional diffusion prior $p_{\theta}(I^e \mid U,R)$ is trained to approximate $p(I^e \mid U,R)$ and generate personalized embeddings:
\begin{equation}
I^e\mid U,R \sim p_{\hat{\theta}}(I^e \mid U,R).
\label{eq:sample_embed}
\end{equation}
Second, a pretrained text-to-image model with an IP-Adapter $p(I \mid I^e)$ approximates $p(I \mid I^e, U,R)$  and decodes these embeddings into visual samples:
\begin{equation}
I\mid I^e\sim p(I \mid I^e). 
\label{eq:decode}
\end{equation}
This approximation assumes that CLIP embeddings retain the information relevant for user preference; prior work shows that CLIP spaces exhibit smooth semantic directions and well-structured manifolds~\cite{radford2021learningtransferablevisualmodels,ramesh2022hierarchicaltextconditionalimagegeneration,levi2025the}, making them a suitable domain for modeling $p(I^e \mid U,R)$. Our key insight is that user preferences, when expressed as implicit feedback, naturally define regions of CLIP-space that can be learned as a conditional diffusion prior. 

\vspace{.1cm} \noindent \textbf{Conditional diffusion prior.}
Let $I^e \in \mathbb{R}^d$ denote a CLIP image embedding. 
We model the conditional prior $p_\theta(I^e \mid U, R)$ via a standard forward diffusion process:
\begin{align}
q(I_t^e \mid I_0^e)
  &= \mathcal{N}\!\left(
      \sqrt{\bar\alpha_t}\, I_0^e,\,
      (1-\bar\alpha_t)\mathbf{I}
    \right),
\end{align}
with $\bar\alpha_t=\prod_{s=1}^{t}\alpha_s$. 
The denoising network $f_\theta$ is trained to predict the clean embedding $I_0^e$ directly from a noisy sample:
\begin{align}
\mathcal{L}_{x_0}(\theta)
  = \mathbb{E}_{I_0^e,\,U,R,\,t,\,\epsilon}\!
    \left[
      \big\|
        f_\theta(I_t^{e},t,U,R)
        - I_0^e
      \big\|_2^2
    \right],
\end{align}
where $\epsilon\!\sim\!\mathcal{N}(0,\mathbf{I})$ and $t\!\sim\!\mathrm{Uniform}\{1,\ldots,T\}$.
At inference, classifier-free guidance (CFG) is applied as
\begin{align}
f_\theta^{(s)} = f_\theta^{(0)} + \omega\!\left(f_\theta^{(1)} - f_\theta^{(0)}\right),
\end{align}
where $f_\theta^{(1)}=f_\theta(I_t^e,t,U,R)$ and $f_\theta^{(0)}=f_\theta(I_t^e,t,\varnothing,\varnothing)$. 
The guidance scale $\omega$ controls personalization strength. 
The final denoised embedding $\hat I_0^e$ is decoded into an image using the frozen decoder, yielding personalized samples consistent with the target user and rating.

\vspace{.1cm} \noindent \textbf{REBECA Prior Architecture.} Our diffusion prior is intentionally lightweight: a $4.4$M-parameter transformer with 6 PriorBlocks (\cref{sup:architecture}), AdaLN-Zero conditioning, and a learned tokenizer that splits CLIP embeddings into tokens. Conditioning enters through user, rating, and timestep tokens combined via a small MLP. The small scale allows end-to-end training in $<10$ minutes on a single RTX-4090, enabling rapid iteration and scalability to large user sets. A full description of our training protocol, and final configuration may be found in \cref{sup:training-protocol}.

\section{\texttt{REBECA} Under Control}
\label{sec:rebeca_under_control}
We design a controlled simulation to evaluate \texttt{REBECA} when user preferences are known.

\subsection{Dataset and Setup}
We construct a controlled setting using dSprites~\cite{dsprites17}, restricting to red/blue hearts and squares (four shape–color pairs). Four synthetic users each prefer one pair $(s_U,c_U)$, and we generate ratings $R \in \{0,1\}$ via a simple probabilistic rule that assigns high probability ($0.95$) to the preferred pair, low probability ($0.05$) to the opposite, and intermediate probability ($0.10$) to mismatched shape or color. We sample $40{,}000$ rated images and split them $90/10$.

\subsection{Image Generation}
\begin{figure}[h]
\centering
\includegraphics[width=0.9\linewidth]{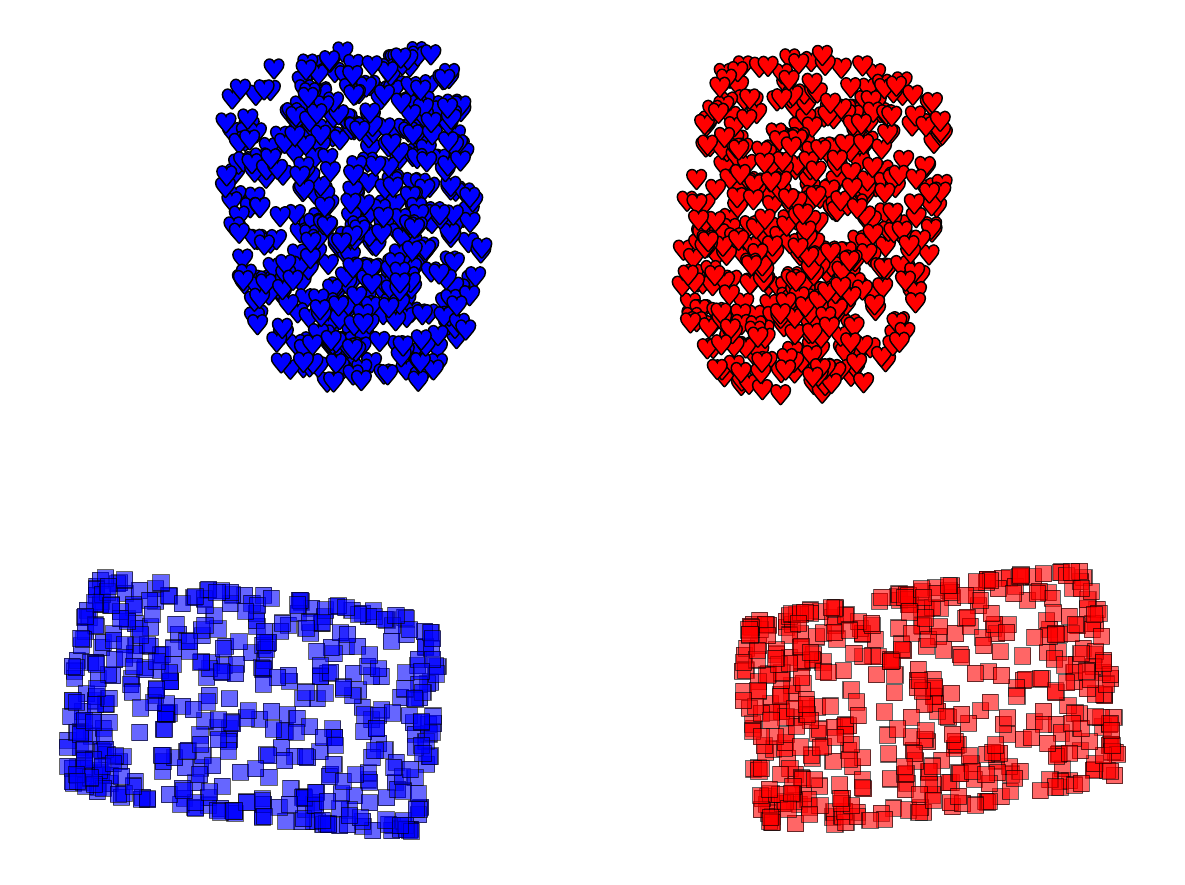}
\caption{UMAP~\cite{umapMcInnes2018} projection of VAE embeddings. Color and shape clusters are cleanly separated.}
\label{fig:umap}
\end{figure}

\texttt{REBECA} is implemented using a lightweight Variational Autoencoder (VAE) \cite{VAE-2019, kingma2022autoencodingvariationalbayes} with a 32-dimensional latent space (Figure~\ref{fig:umap}) trained on the dSprites dataset. We construct a small VAE solely to define a ground-truth latent space with a frozen decoder that reconstructs final images from embeddings.. The encoder provides compact representations of dSprites, and the learned prior generates new embeddings aligned with user taste. 

\subsection{Evaluation}
\label{subsec:evaluation}
Evaluation proceeds in two stages.  
(1)~\textbf{Qualitative:} UMAP projections visualize whether user embeddings cluster coherently according to preferences.  
(2)~\textbf{Quantitative:} \texttt{REBECA} is compared with (a) a mean-embedding baseline that generates a single prototype image per user and (b) a random baseline sampling uniformly from training data.  
For each user, \texttt{REBECA} and the random baseline generate $k{\in}\{1,5,10,20,25\}$ samples.  
We compare REBECA to (a) a mean-embedding baseline that generates a single prototype per user, and (b) a random baseline that samples uniformly from the training set. For $k\in\{1,5,10,20\}$, we report Precision@k (fraction of generated images whose nearest neighbor in the test set belongs to the user’s liked subset) and Recall@k (fraction of liked test images for which at least one generated sample is the nearest neighbor).

\begin{table}[t]
\centering
\small
\caption{\textbf{Precision@k and Recall@k} averaged over users in the controlled simulation. 
Recall is scaled by $\times 10^{-3}$. 
The mean baseline is deterministic and thus lacks diversity.}
\label{tab:pr}
\begin{tabular}{l c c c}
\toprule
Method & $k$ & Precision@$k$ & Recall@$k$ ($\times 10^{-3}$) \\
\midrule
\multirow{5}{*}{\texttt{REBECA}}
 & 1  & 0.782  & 0.167  \\
 & 5  & 0.782  & 0.835  \\
 & 10 & 0.789  & 1.685  \\
 & 20 & 0.799  & 3.409  \\
\midrule
\multirow{1}{*}{Mean (deterministic)} 
 & 1  & \textbf{1.000}  & 0.214 \\
\midrule
\multirow{5}{*}{Random}
 & 1  & 0.290  & 0.062  \\
 & 5  & 0.282  & 0.301  \\
 & 10 & 0.287  & 0.613  \\
 & 20 & 0.291  & 1.244  \\
\bottomrule
\end{tabular}
\end{table}

% Mirror the TOP half horizontally (left↔right) without any arithmetic
\newcommand{\MirrorTopHoriz}[2][\linewidth]{%
  \begin{tikzpicture}
    % base image
    \node[anchor=south west,inner sep=0] (base) {\includegraphics[width=#1]{#2}};
    % clip to the top half using midpoint along the left edge
    \begin{scope}
      % Midpoint syntax: (A)!t!(B) = point t along segment AB.
      % Left-edge halfway point is between south west and north west:
      \clip ($(base.south west)!0.44!(base.north west)$) rectangle (base.north east);
      % draw a horizontally mirrored copy over the clipped region
      \node[anchor=south west,inner sep=0] at (base.south west)
        {\reflectbox{\includegraphics[width=#1]{#2}}};
    \end{scope}
  \end{tikzpicture}%
}

\begin{figure*}[t]
\centering

\newlength\labelw
\setlength{\labelw}{0.01\linewidth}
\newlength\imgw
\setlength{\imgw}{0.24\linewidth}

\setlength{\tabcolsep}{1pt}
\setlength{\arrayrulewidth}{0.25pt}
\arrayrulecolor{black!100}

\begin{tabular}{@{}p{\labelw} |c|c|c|c|@{}}
   & \scriptsize\bfseries User $0$ & \scriptsize\bfseries User $1$ & \scriptsize\bfseries User $2$ & \scriptsize\bfseries User $3$ \\[0.3em]
  \toprule
  \raisebox{2.5\height}{\rotatebox{90}{\scriptsize\bfseries Likes}} &
  \includegraphics[width=\imgw]{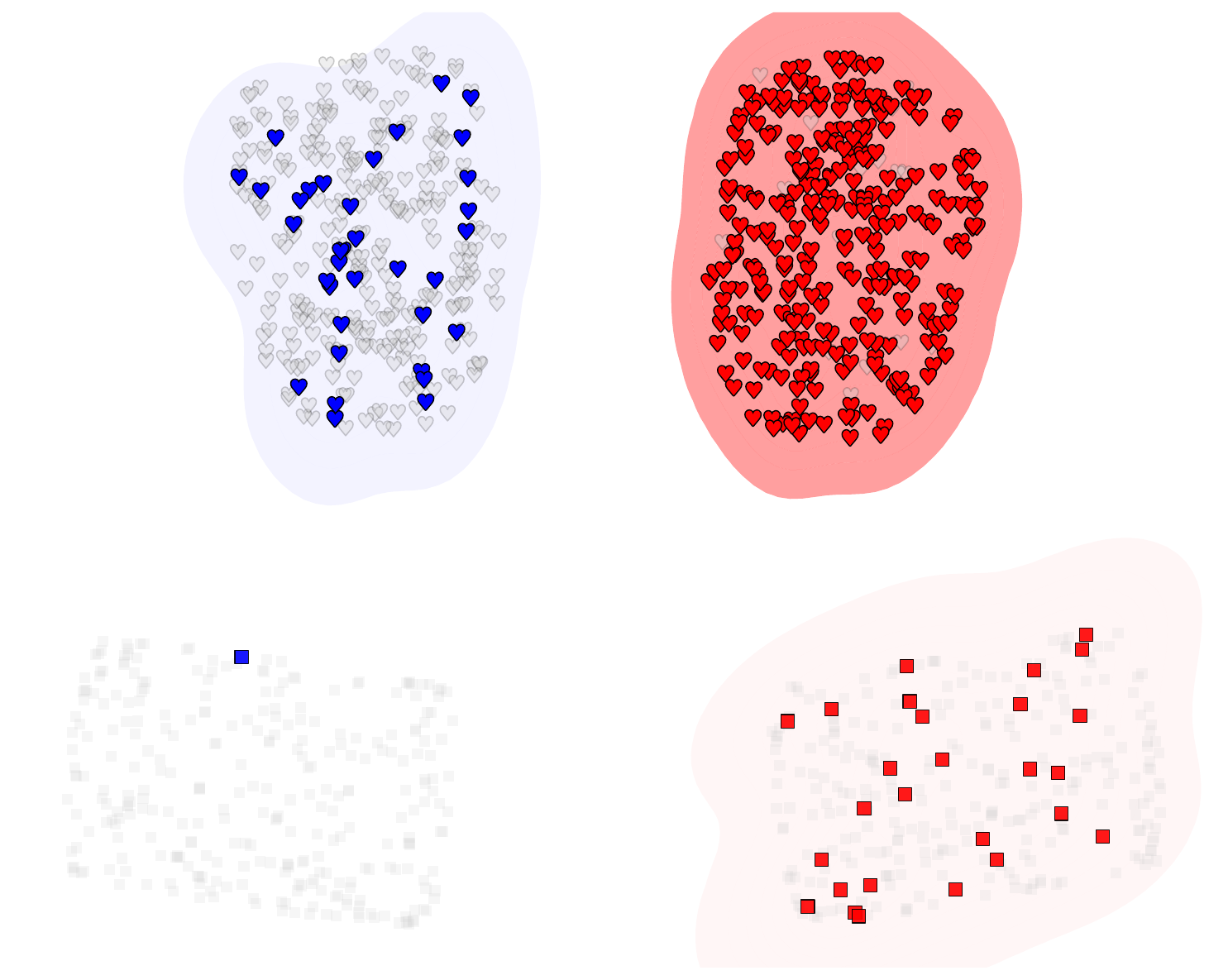} &
  \includegraphics[width=\imgw]{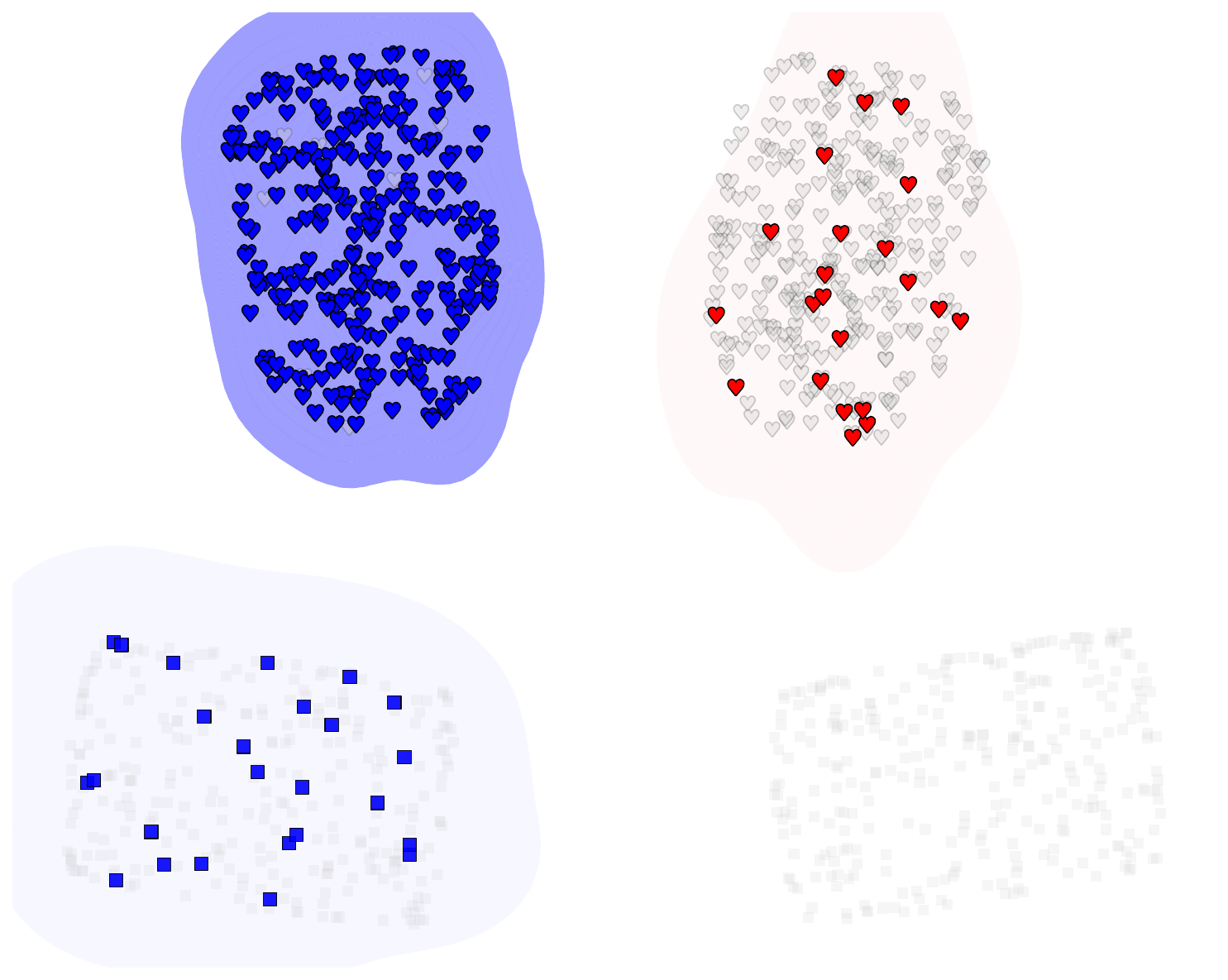} &
  \includegraphics[width=\imgw]{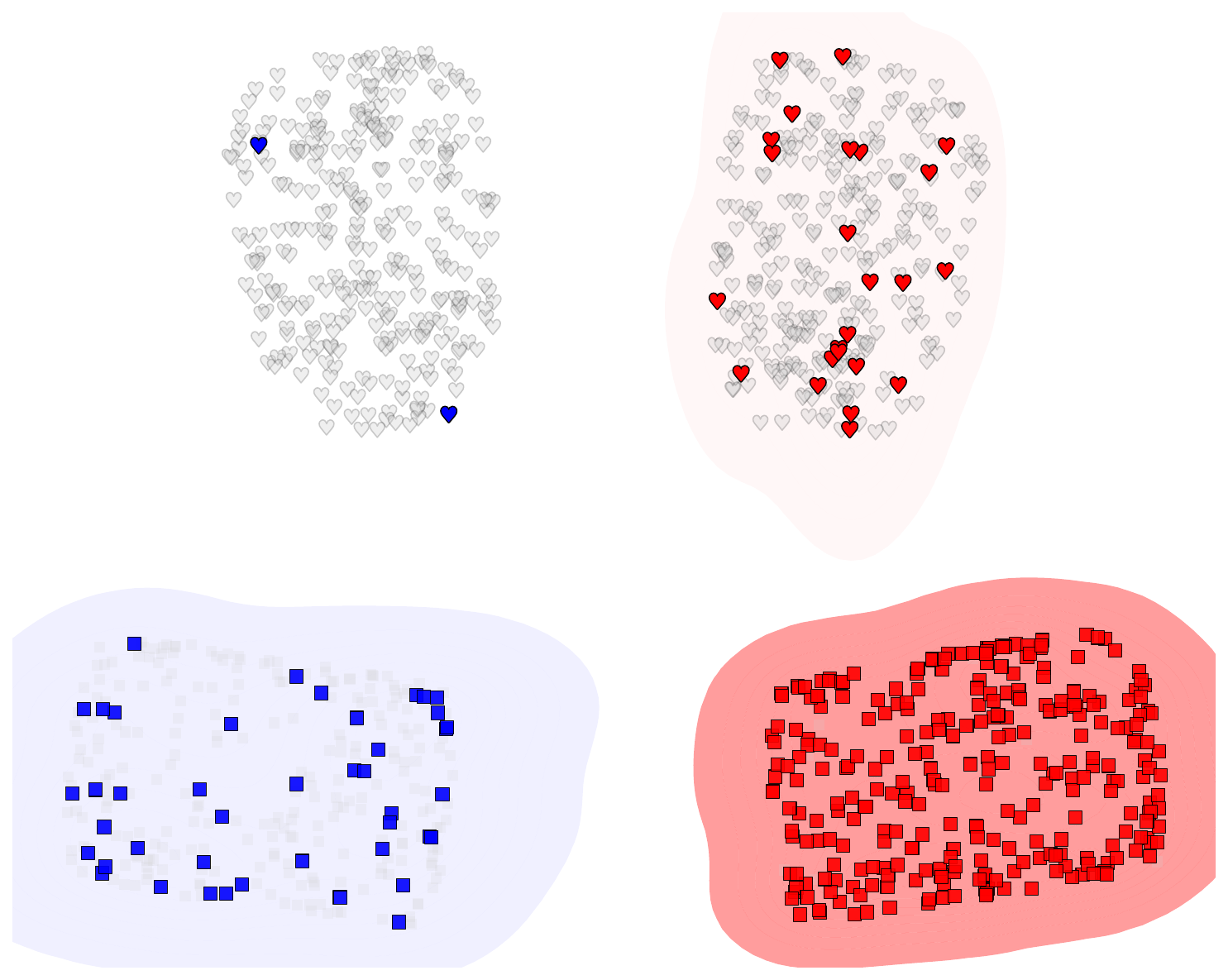} &
  \includegraphics[width=\imgw]{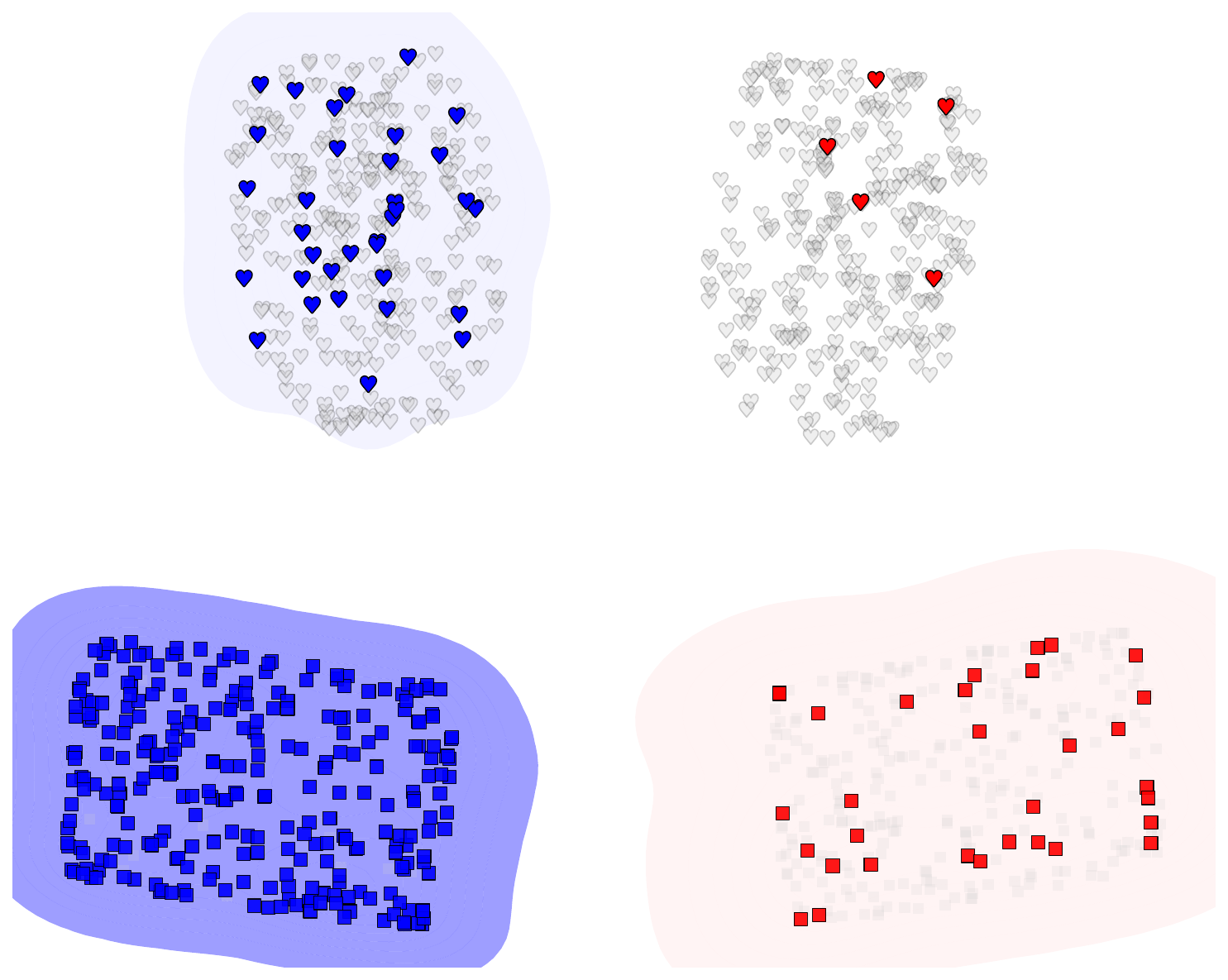} \\[-0.3em]
  \hline
  \raisebox{0.75\height}{\rotatebox{90}{\scriptsize\bfseries Generations}} &
  \MirrorTopHoriz[\imgw]{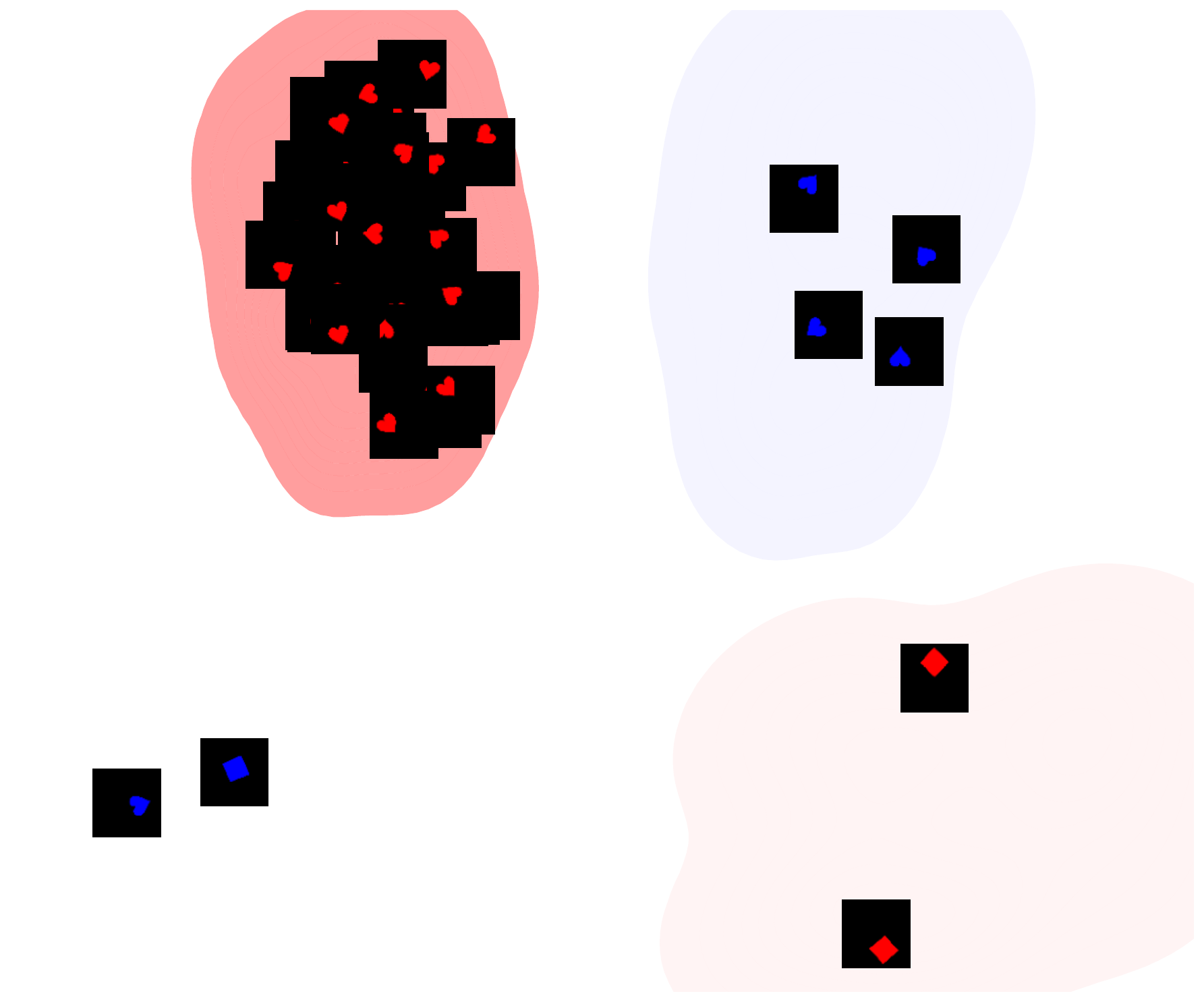} &
  \MirrorTopHoriz[\imgw]{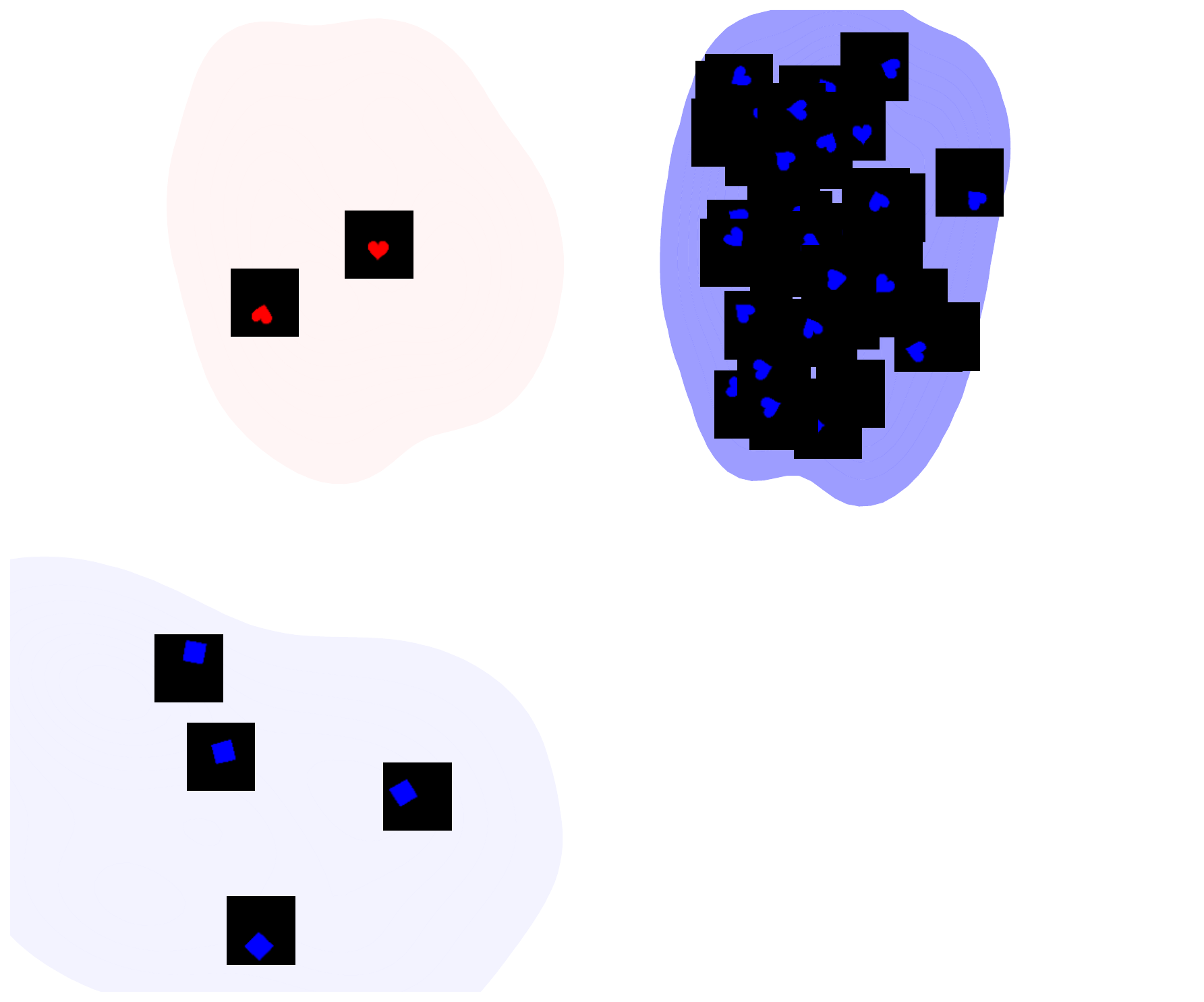} &
  \MirrorTopHoriz[\imgw]{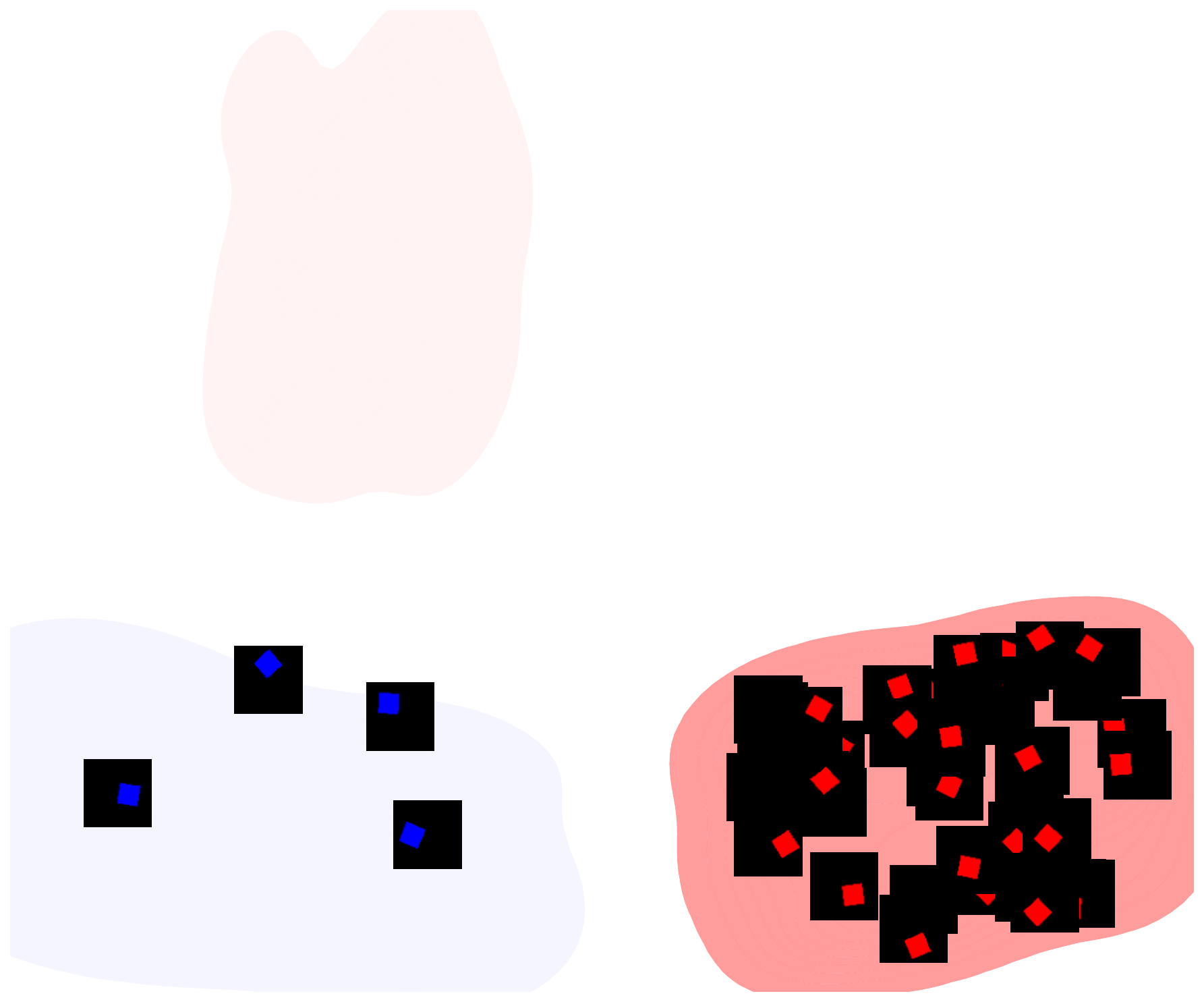} &
  \MirrorTopHoriz[\imgw]{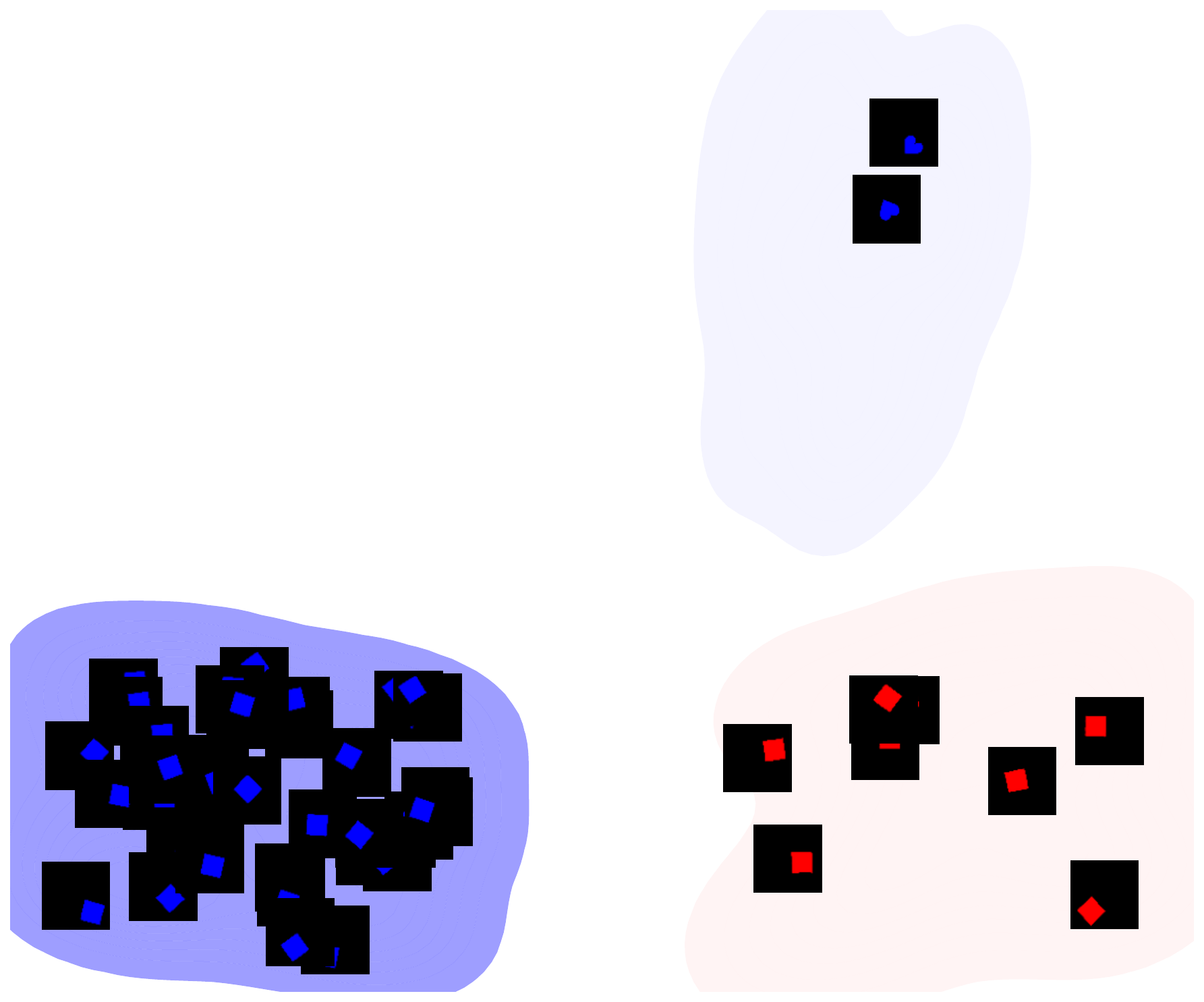}\\
  \bottomrule
\end{tabular}

\caption{Per-user visualization in the controlled setting.  
Top: liked samples for each user.  
Bottom: images generated by \texttt{REBECA} using the frozen VAE decoder.  
\texttt{REBECA} captures each user's preference manifold while maintaining diversity.}
\label{fig:users-by-type}
\end{figure*}

\subsection{Results}  
Figure~\ref{fig:users-by-type} shows that generated samples align with users’ liked regions in latent space.  
Table~\ref{tab:pr} confirms that \texttt{REBECA} achieves the best precision-recall balance, capturing user tastes while preserving diversity.  

This controlled setting validates that REBECA recovers user-specific preference manifolds when ground truth is known. In Section ~\cref{sec:exp} we show that the same structure appears in real-world behavioral data, using a learned verifier as a surrogate for unknown preferences.

\section{\texttt{REBECA} In The Wild}\label{sec:exp}
In this section, we evaluate \texttt{REBECA} against strong baselines in a realistic scenario, using images rated by real users.
\begin{figure}
    \centering
    \includegraphics[width=0.9\linewidth]{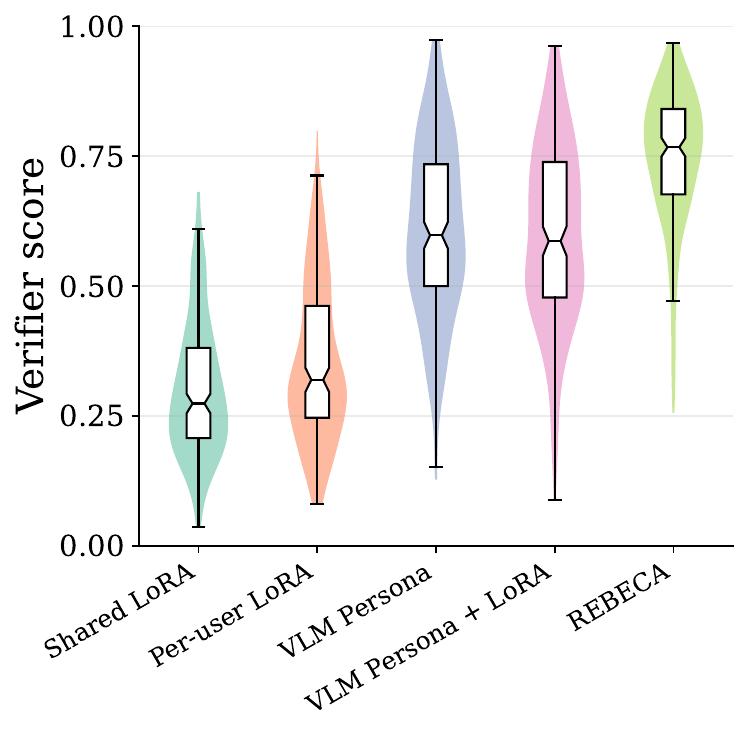}
    \caption{Comparison of personalization performance across generation approaches. \texttt{REBECA} achieves the highest user scores, surpassing VLM-based baselines and LoRA fine-tuning variants.}
    \label{fig:personalization_results}
\end{figure}
\subsection{Dataset and Setup}
\label{subsec:rebeca-data}
We employ the FLICKR-AES dataset \cite{Ren_2017_ICCV}. The dataset contains a curated sample of $40,988$ Creative Commons-licensed photos from FLICKR, rated for aesthetic quality on a scale of 1 to 5. Annotations were collected using Amazon Mechanical Turk, with each image receiving ratings from five different workers. In total, 210 workers contributed, scoring 40,988 images for a total of $193208$ ratings. 

% Initial setup and transformations
Thus, our data consist of triplets of users, ratings, and images $(U, R, I)$. CLIP embeddings $I^e$ are deterministic mappings of images, and we obtain them with a single pass through the appropriate CLIP \textit{ViT-H-14} \cite{cherti2023reproducible} encoder. We further map the ratings to binary ratings, representing like and dislike: given a pair of user and a rated image, we let $R=1$ if the rating is $4$ or higher, and $R=0$ otherwise. At the training stage, we fit our image embedding prior with $(U, R, I^e)$. Importantly, image captions are not part of our data, and we uniquely rely on users' historical ratings. 
\subsection{Image Generation}
%We restate our objective as enhancing catalogs in accordance with users' preferences. This means that at the sampling stage, we generate image embeddings that are personalized based on the user's historical interactions.
To ensure that the sampled embeddings align with user preferences, we set $R=1$ to indicate a strong preference signal, thereby prioritizing images that align with previously liked content. At first sight, it might appear that training on \textit{liked} and \textit{disliked} images to generate only positive samples is unnecessary. However, this decision is principled, and the reasoning is twofold. First, learning from both positive and negative feedback enables the model to build more informative user embeddings, capturing a fuller picture of individual preferences. Second, there is strong empirical evidence that conditional generative models consistently outperform their unconditional counterparts in terms of sample quality and alignment \cite{dhariwal2021diffusionmodelsbeatgans, brock2019largescalegantraining}.

\subsection{Evaluation}
All image generation uses \emph{Stable Diffusion v1.5}~\cite{podell2023sdxlimprovinglatentdiffusion,rombach2021highresolution} as a frozen decoder, operating at $512{\times}512$ resolution in fp16. We generate $25$ images per user using $50$ denoising steps with classifier-free guidance (CFG) set to $5$, balancing personalization and diversity. We used empty system prompts during these experiments. The diffusion prior outputs a personalized embedding $I^e$, which is then provided to the SD\,1.5 pipeline equipped with the IP-Adapter~\cite{ye2023ip-adapter}.

\subsubsection{Baselines}
\textit{Baselines require image captions.} To obtain semantic descriptions of the training images, we use the \textit{LLaVA-1.5-7B} multimodal model~\cite{liu_llava} as an image-language tagger. Each image is passed through LLaVA with a fixed JSON-style instruction that requests a concise caption together with short lists of objects, attributes, styles, and colors. See \cref{sup:image-tagging} for a more detailed specification. The model outputs are parsed into a structured table of captions and visual keywords used downstream for preference modeling. Next, we introduce the baselines in more detail.

\vspace{.1cm} \noindent \textbf{LoRA per User.}
 We train a separate LoRA adapter~\cite{hu2021loralowrankadaptationlarge} for each user on that user’s Flickr subset, keeping the CLIP text encoder frozen. Hyperparameters (rank, steps, warmup) are adapted to the number of images per user (\cref{suppl:lpu}).

 \vspace{.1cm} \noindent \textbf{Shared LoRA.}
A single LoRA adapter is trained on the union of all users, then loaded into a fresh SD1.5 pipeline at evaluation time. See \cref{suppl:shared-lora}.

\vspace{.1cm} \noindent \textbf{VLM Personas.}
Following PMG-style text conditioning~\cite{shen2024pmg}, we derive per-user textual personas from LLaVA-generated tags, including short descriptions and positive/negative keyword lists. Prompts are built by combining persona text with sampled positive keywords and standard degradation terms in the negative prompt. We refer the reader to \cref{sup:vlm-persona} for a comprehensive description.
% \texttt{\{user\_id, persona, keywords\_positive, keywords\_negative\}}. Prompts are composed by cycling $1$-$2$ positive keywords plus persona style; when \texttt{persona} is missing we append a neutral stylistic suffix (``high quality, detailed, natural lighting''). Negative prompts include standard degradation terms (``blurry, deformed, overexposed'').

\vspace{.1cm} \noindent \textbf{VLM Personas + LoRA.}
We also combine per-user LoRA with VLM personas, loading each user’s LoRA adapter and prompting with their persona during generation.

\begin{figure*}[t]
\centering
\begin{tabular}{cc} \includegraphics[width=0.49\linewidth]{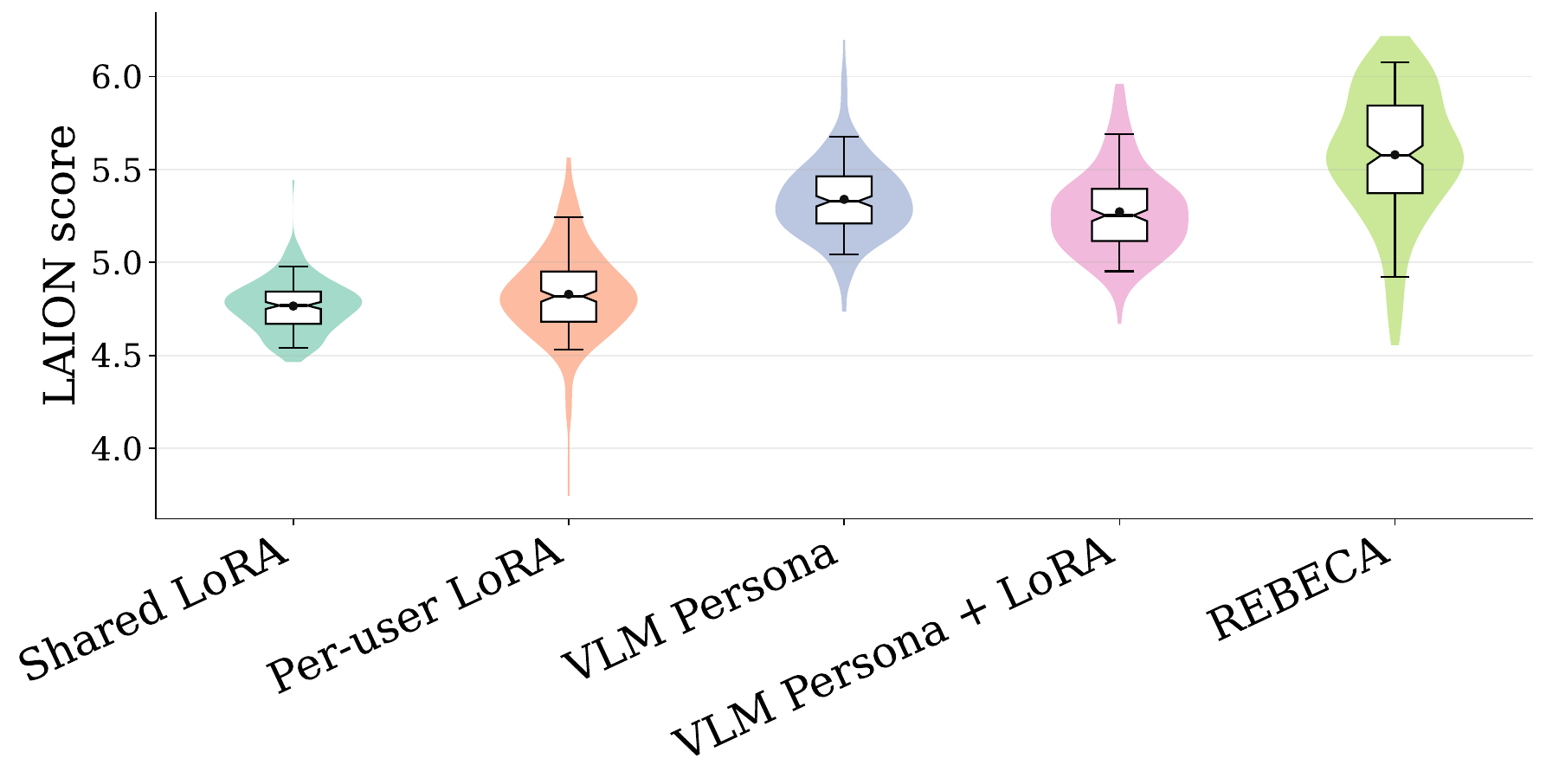} & \includegraphics[width=0.49\linewidth]{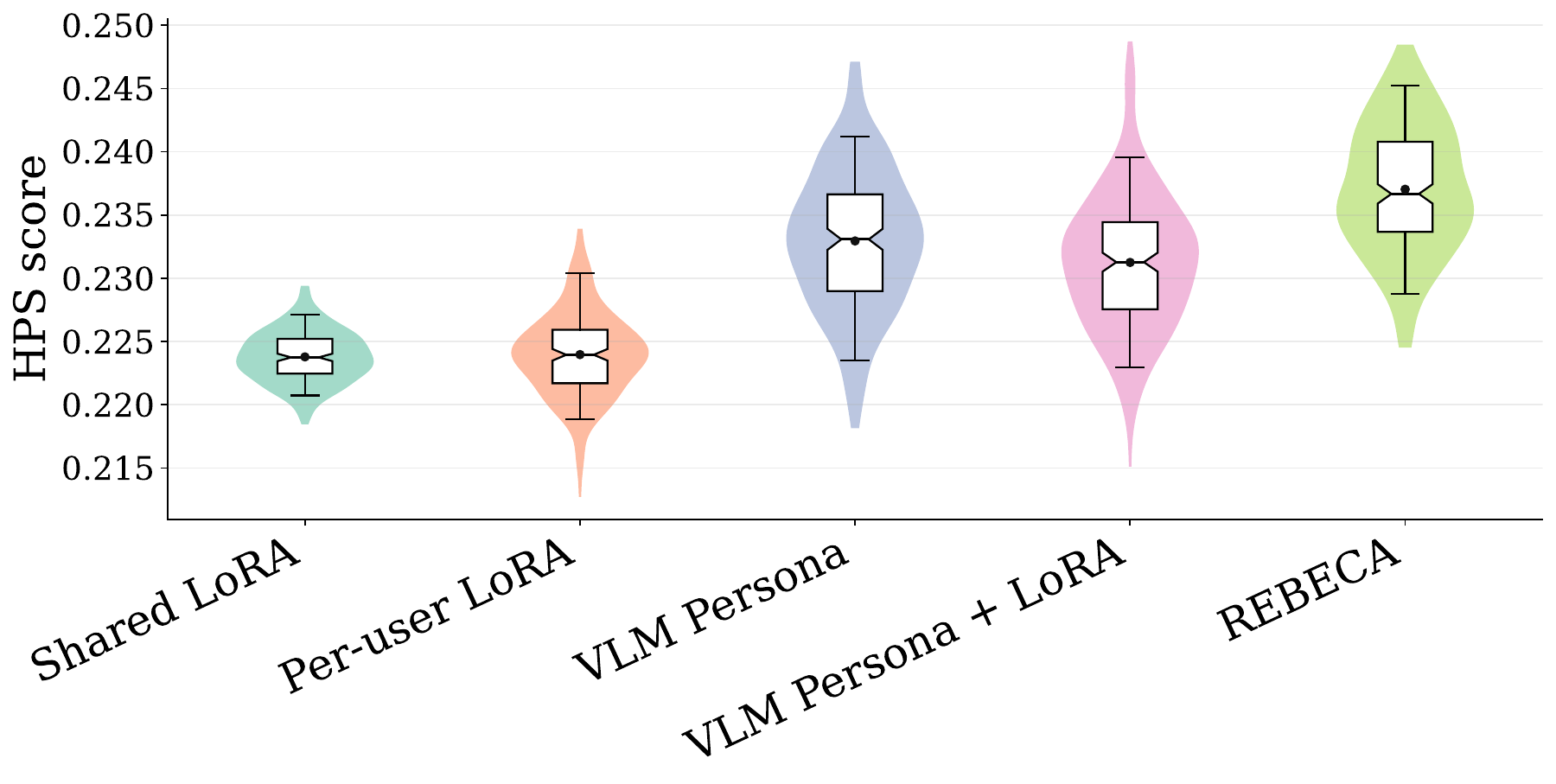} \\ (a) LAION aesthetic predictor & (b) HPSv2 overall quality predictor \end{tabular}
\caption{
Quantitative comparison of aesthetic and quality predictors across personalization methods.
(a) The LAION aesthetic predictor captures visual appeal independently of textual prompts.
(b) The HPSv2 predictor measures prompt-conditioned human preference and image quality.
Across both metrics, \texttt{REBECA} consistently achieves the best performance.
}
\label{fig:aesthetic_predictors}
\end{figure*}
\subsubsection{Metrics}
In addition to precision@$k$ and recall@$k$ (Section \ref{subsec:evaluation}), we introduce additional metrics for quantitative evaluation.

\vspace{.1cm} \noindent \textbf{Personalization Verifier.} Since we cannot directly evaluate the preferences of users from the FLICKR-AES dataset on generated images, we cannot immediately determine whether the images produced by \texttt{REBECA} align with individual tastes. To address this, we train a predictive model that estimates the probability that a particular user $U$ will like a given image $I$. Models serving this kind of role are often referred to as \emph{verifiers} and are increasingly common in the literature on generative modeling \citep{cobbe2021training,lightman2023let}.

Our verifier is based on the \emph{Neural matrix factorization model} \citep{he2017neural}. Specifically, the verifier $v(U,I)$ approximates $\mathbb{P}(R{=}1 \mid U,I)$ and is given by
\begin{equation}
v(U,I)
= g_{\gamma}\big(U,\texttt{CLIP}(I)\big),
\label{eq:verifier}
\end{equation}
where $g_\gamma$ is a neural network, each user index $U$ is mapped to a trainable embedding vector, and $\texttt{CLIP}(\cdot)$ represents a fixed image embedding model\footnote{We use \texttt{OpenCLIP-ViT-bigG-14}~\cite{cherti2023reproducible}, a different CLIP backbone than in Section \cref{subsec:rebeca-data}, to avoid representation leak.}. The verifier is trained by minimizing the binary cross-entropy loss\footnote{This configuration achieves an ROC-AUC of approximately $87\%$ on both the training and test sets, with roughly $30\%$ of images in class 1.} using the same training data employed to train the baselines and \texttt{REBECA}\footnote{We use the training set to fit the verifier because it provides a larger number of labeled examples, allowing for a more accurate estimation of user preferences. While this could, in principle, introduce bias if both the verifier overfits, we find this unlikely in our case. As shown in Appendix~\ref{sec:diag}, the difference in the verifier's performance on the training and test sets is not statistically different.}. We then use the trained verifier $\hat{v}$ to quantify the personalization quality of generative models. Specifically, we define the personalization score of a generative model $\hat{p}(I \mid U, R=1)$ for user $U$ as
\begin{equation}
\text{Score}\!\left(\hat{p}(I\mid U,R{=}1)\right)
= \mathbb{E}_{\hat{p}(I\mid U,R{=}1)}[\,\hat{v}(U,I)\,],
\label{eq:expected_score}
\end{equation}
which we estimate empirically by sampling generated images for user $U$. This allows us to quantitatively compare different variants of \texttt{REBECA} and measure their ability to generate user-personalized images.

\vspace{.1cm} \noindent \textbf{Aesthetics and Overall Quality.} Personalization methods should preserve key aspects of image quality, including aesthetics and overall visual appeal. To evaluate whether \texttt{REBECA} and the baselines maintain high-quality generation while achieving personalization, we employ the LAION aesthetic predictor \citep{laion_aesthetic_predictor} and the Human Preference Score (HPSv2) \citep{wu2023human} to assess image aesthetics and perceptual quality. The two methods produce scalar scores that can be used to compare the quality of generated images.

\subsection{Results}

\begin{figure*}[t]
  \centering
  \setlength{\tabcolsep}{1pt}
  \renewcommand{\arraystretch}{0.90}

  \newcommand{\interpheader}{
    \makebox[0.19\linewidth][c]{Original} &
    \makebox[0.19\linewidth][c]{$t=0.1$} &
    \makebox[0.19\linewidth][c]{$t=0.2$} &
    \makebox[0.19\linewidth][c]{$t=0.3$} &
    \makebox[0.19\linewidth][c]{$t=0.4$}
  }

  \begin{tabular}{@{}lccccc@{}}
    & \interpheader \\[3pt]

    % ---------------- User 0 ----------------
    % \rotatebox{90}{\scriptsize User 0} &
    % \imgscoredisliked{figures/editing/user_0/dog_t00.jpg}{0.71} &
    % \imgscore{figures/editing/user_0/dog_t01.png}{0.81} &
    % \imgscore{figures/editing/user_0/dog_t02.png}{0.79} &
    % \imgscore{figures/editing/user_0/dog_t03.png}{0.76} &
    % \imgscore{figures/editing/user_0/dog_t04.png}{0.74} \\[3pt]

    % ---------------- User 2 ----------------
    \rotatebox{90}{\scriptsize User 2} &
    \imgscoredisliked[trim={0 0 0 4.5cm},clip]{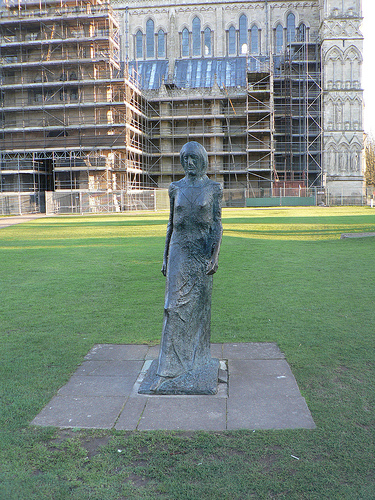}{0.16} &
    \imgscore{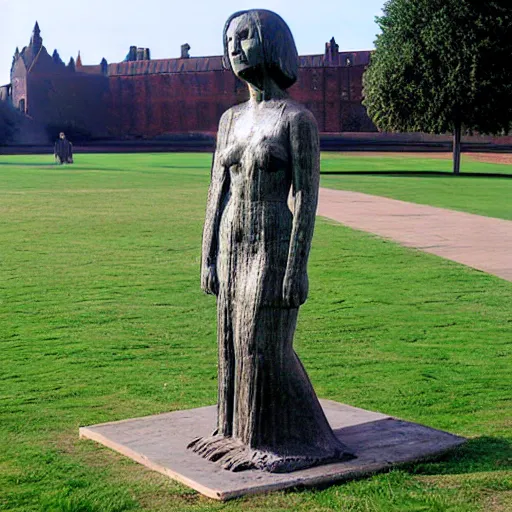}{0.56} &
    \imgscore{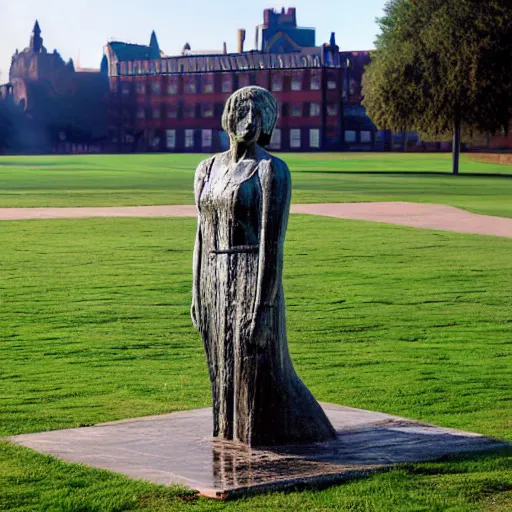}{0.70} &
    \imgscore{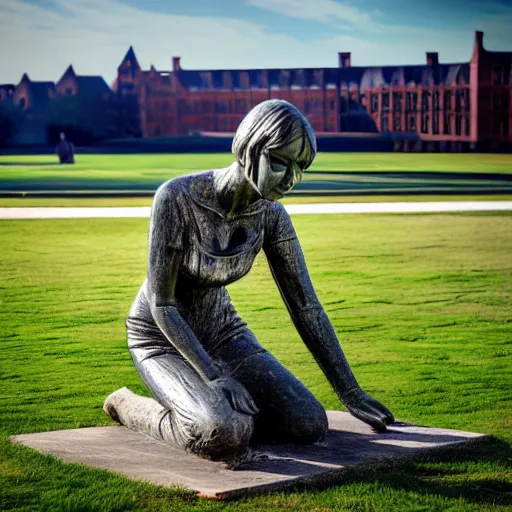}{0.66} &
    \imgscore{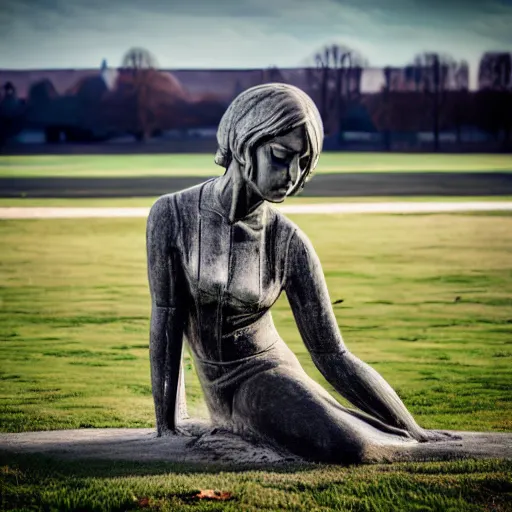}{0.75} \\[3pt]

    % ---------------- User 12 ----------------
    % \rotatebox{90}{\scriptsize User 12} &
    % \imgscoredisliked{figures/editing/user_12/bus_t00.jpg}{0.58} &
    % \imgscore{figures/editing/user_12/bus_t01.png}{0.59} &
    % \imgscore{figures/editing/user_12/bus_t02.png}{0.71} &
    % \imgscore{figures/editing/user_12/bus_t03.png}{0.72} &
    % \imgscore{figures/editing/user_12/bus_t04.png}{0.73} \\[3pt]

    % ---------------- User 21 (your new row) ----------------
    \rotatebox{90}{\scriptsize User 21} &
    \imgscoredisliked{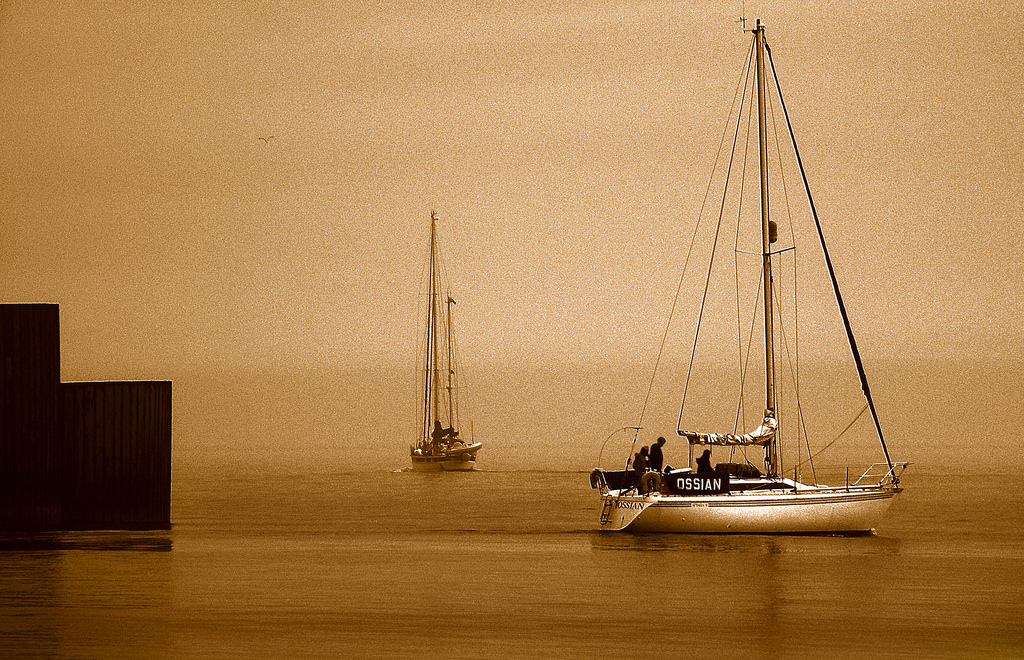}{0.35} &
    \imgscore{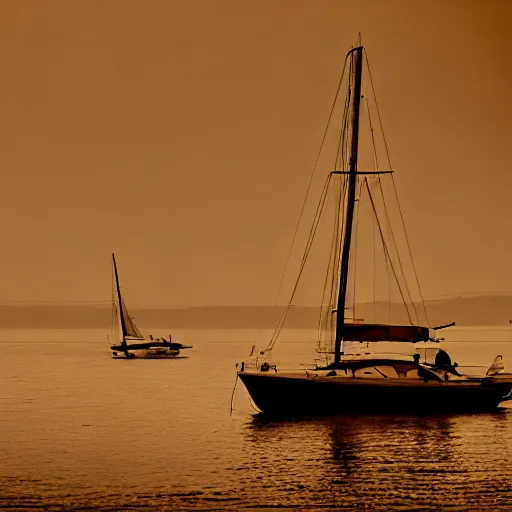}{0.51} &
    \imgscore{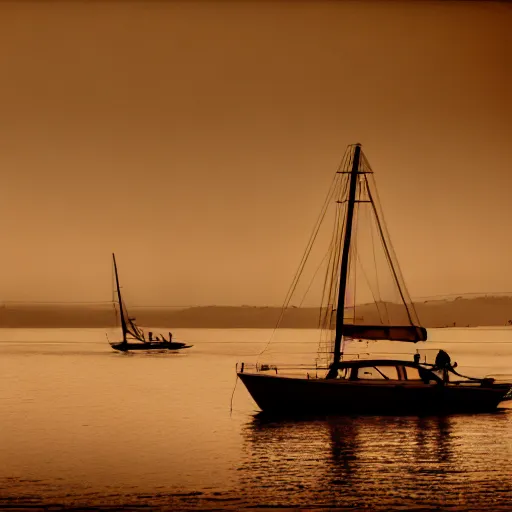}{0.57} &
    \imgscore{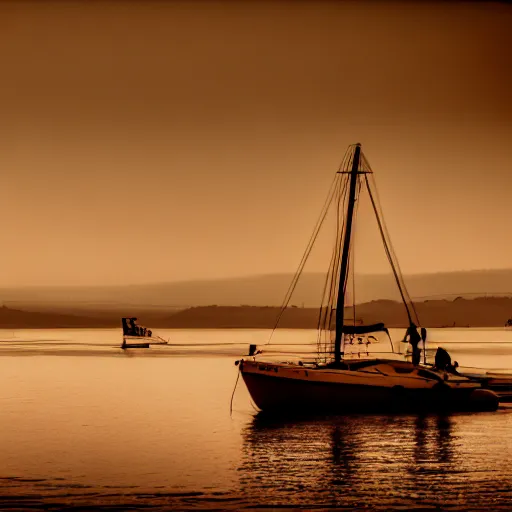}{0.66} &
    \imgscore{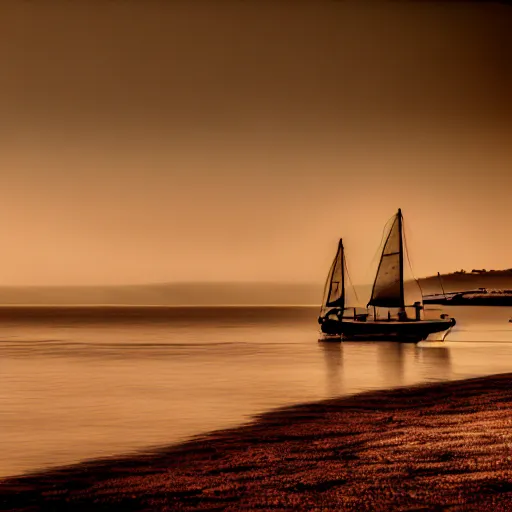}{0.74} \\
  \end{tabular}

  \vspace{-0.3em}
  \caption{Interpolation results between latent representations (rows: users 2, 21).
  Leftmost image in each row was originally \emph{disliked} (red badge).
  Columns show increasing $t\!\in\!\{0.1,0.2,0.3,0.4\}$ from the original.
  Overlay badges show the predicted preference score.}
  \label{fig:interpolations}
\end{figure*}
\vspace{.1cm} \noindent \textbf{Precision@$k$ and Recall@$k$.}
Table~\ref{tab:pr-at-k} reports macro-averaged precision, recall, and F1 for 210 users.
Across both @1 and @5, \texttt{REBECA} matches or surpasses all baselines.
At lower prior CFG values the model attains the highest recall, sampling broadly from each user's preference manifold but at lower precision.
As CFG increases, generation concentrates around dominant preferences, raising precision while slightly reducing recall.
Macro-F1 remains highest for \texttt{REBECA}, confirming the strongest overall precision-recall balance.

\begin{table}[h]
\centering
\small
\setlength{\tabcolsep}{2.2pt}
\begin{tabular}{lcccccc}
\toprule
 & \multicolumn{3}{c}{@1} & \multicolumn{3}{c}{@5} \\
\cmidrule(lr){2-4}\cmidrule(lr){5-7}
Model & P & R & F1 & P & R & F1 \\
\midrule
Shared LoRA            & 0.455 & 0.718 & 0.540 & 0.471 & 0.991 & 0.623 \\
Per-user LoRA          & 0.485 & 0.713 & 0.557 & 0.470 & 0.986 & 0.621 \\
VLM Persona            & 0.507 & 0.685 & 0.556 & 0.472 & 0.983 & 0.621 \\
VLM Persona + LoRA     & 0.511 & 0.673 & 0.555 & 0.475 & 0.986 & 0.626 \\
\texttt{REBECA} (CFG=$3.0$)             & 0.524 & \textbf{0.767} & \textbf{0.605} & 0.478 & \textbf{0.997} & \underline{0.630} \\
\texttt{REBECA} (CFG=$5.0$)              & 0.556 & \underline{0.714} & \textbf{0.605} & 0.478 & \underline{0.989} & 0.629 \\
\texttt{REBECA} (CFG=$7.0$)             & \underline{0.579} & 0.647 & \underline{0.584} & \underline{0.480} & 0.976 & 0.628 \\
\texttt{REBECA} (CFG=$9.0$)             & \textbf{0.605} & 0.610 & 0.575 & \textbf{0.484} & 0.972 & \textbf{0.631} \\
\bottomrule
\end{tabular}
\caption{Macro-averaged precision/recall/F1 over 210 users for retrieval of ground-truth likes from a liked+disliked pool using cosine similarity. Bold indicates best, underline indicates second-best per column.}
\label{tab:pr-at-k}
\end{table}

\vspace{.1cm} \noindent \textbf{Personalization.} We compare its performance against several baseline methods. Figure~\ref{fig:personalization_results} presents box-plots showing the distribution of user scores for each generation approach. \texttt{REBECA} clearly outperforms all alternatives, with VLM-based methods serving as strong baselines. We also observe that training a separate LoRA per user improves personalization compared to using a shared LoRA across users, though the gain is modest. This suggests that while LoRA introduces efficient low-rank adaptation, it lacks the inductive biases embedded in \texttt{REBECA} that are crucial for personalized image generation. In Appendix \ref{sec:personalization}, we further show disaggregated results.

\vspace{.1cm} \noindent \textbf{Aesthetics and Overall Quality.} Figure \ref{fig:aesthetic_predictors} reports results for both the LAION aesthetic predictor (left) and the HPSv2 human preference metric (right). The relative performance across methods closely mirrors what we observed in Figure \ref{fig:personalization_results}, with \texttt{REBECA} consistently achieving the highest scores. Note that the HPSv2 metric is prompt-dependent: it evaluates the alignment between textual input and generated images. In our main evaluation (\cref{fig:aesthetic_predictors}), we use the generic prompt to capture overall quality ``Realistic image, finely detailed, with balanced composition and harmonious elements.'' In  \cref{sec:aes-quality}, we further evaluate with an empty prompt. In that setting, all generative methods exhibit similar overall quality, indicating that even in the absence of explicit textual guidance, \texttt{REBECA} maintains high image quality while delivering effective personalization.

\begin{figure}[t]
  \centering
  \includegraphics[width=\linewidth, trim={0.2cm 0 0.2cm 0.3cm},clip]{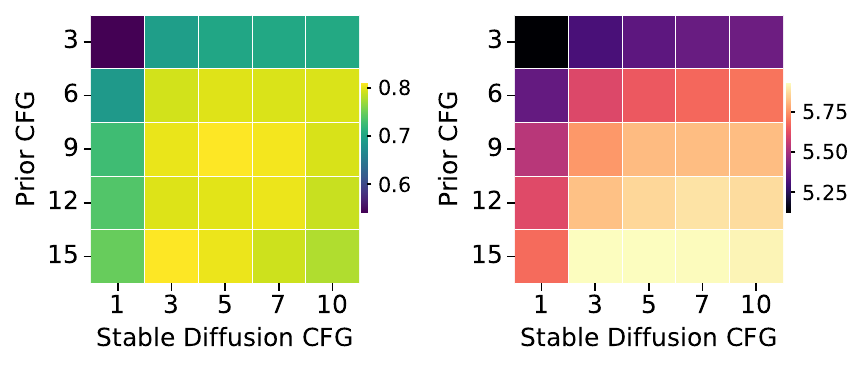}
  \vspace{-0.5em}
  \caption{User-averaged scores by SD$\times$Prior CFGs. Left: verifier score. Right: user-averaged LAION aesthetic score. 
  Higher is better; color scales differ per metric.}
  \label{fig:ablations-medians}
\end{figure}
\vspace{.1cm} \noindent \textbf{Ablations.}
\label{subsec:ablations}
During generation, inference-time parameters such as classifier-free guidance (CFG) and system prompts may, in principle, influence the final output. We therefore analyze \texttt{REBECA} along two axes.

\textbf{(i) Dual guidance.}  
A joint sweep over the diffusion-prior CFG and the Stable Diffusion CFG (Fig.~\ref{fig:ablations-medians}) reveals a clear trade-off between personalization strength and image fidelity. Increasing the prior CFG amplifies user-specific signal, while higher SD CFG improves aesthetic quality but mildly attenuates personalization. This interaction motivates our choice of moderate CFG values in the main experiments.

\begin{figure}[t]
  \centering
  \setlength{\tabcolsep}{1pt}
  \begin{tabular}{@{}cc@{}}
  \includegraphics[width=0.9\linewidth, trim={0.3cm 0 0.2cm 0.3cm},clip]{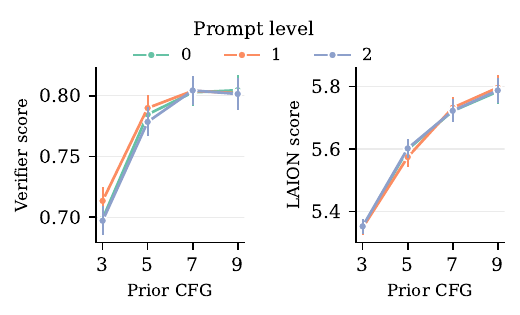} &
  \end{tabular}
  \vspace{-0.4em}
  \caption{Prompt level for (a) verifier score and (b) LAION aesthetic quality. With \texttt{REBECA}, we do not observe a significant difference in scores with changing prompts
  }
  \label{fig:ablations-cross-prompts}
\end{figure}

\textbf{(ii) Prompt control.}  
We further test whether prompt engineering can enhance output quality by evaluating three increasingly structured system prompts (see \cref{sup:system-prompts}). As shown in Fig.~\ref{fig:ablations-cross-prompts}, all prompt levels yield nearly identical verifier and LAION scores across CFG settings, indicating that \texttt{REBECA}’s personalization is effectively \emph{prompt-independent}. Even strong descriptive and negative prompts do not modulate the personalization signal, which instead resides in the latent user-conditioned embedding distribution learned by the diffusion prior, not in the text-conditioning pathway.

\subsection{\texttt{REBECA} generates personalized images}

\begin{figure}[t]
  \centering
  \setlength{\tabcolsep}{1pt}
  \begin{tabular}{@{}cc@{}}
  \includegraphics[width=0.49\linewidth]{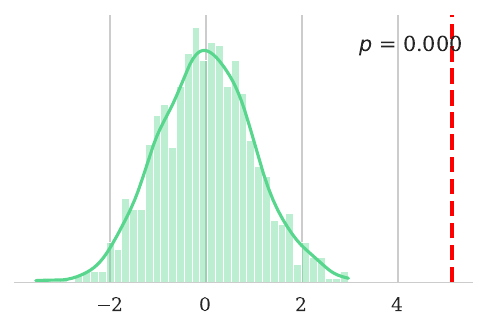} &
    \includegraphics[width=0.49\linewidth]{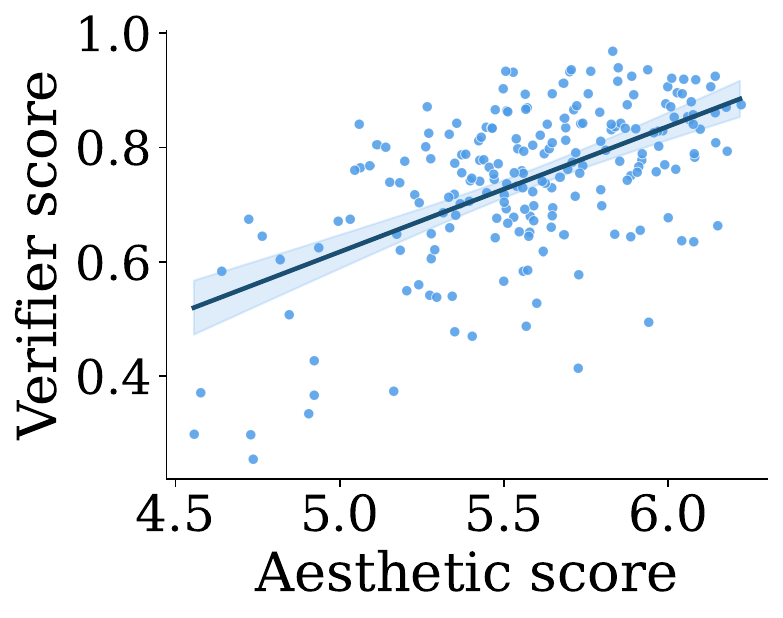} 
  \end{tabular}
  \vspace{-0.4em}
  \caption{(a) Permutation test: correctly matched user-image pairs score nearly five standard deviations above random, indicating genuine personalization beyond aesthetics. (b) Aesthetic quality vs. verifier score.}
  \label{fig:aes-perm-side}
\end{figure}

We test whether \texttt{REBECA}'s gains are due solely to aesthetic quality by measuring personalization through a user-image matching experiment (\cref{fig:aes-perm-side}). First, for each user, we compare the median verifier score of their own images to scores when images are randomly reassigned. The difference (correct minus random) quantifies personalized alignment. A permutation test\footnote{Check the used algorithm in Appendix \ref{sec:hyp-test}.} \citep{lehmann1986testing} of $H_0: U \ind I$ shows that correct matches score nearly five standard deviations above random (p $< 10^{-3}$, $\alpha=0.05$), confirming strong user-specific preference capture. Second, correlating verifier scores with average LAION aesthetic scores reveals a significant but modest relationship ($R^2 = 0.32$), indicating that aesthetic quality contributes partially to satisfaction but does not fully explain \texttt{REBECA}'s personalization. 

% \begin{figure}[h]
% \centering
% \includegraphics[width=.9\linewidth]{figures/aesthetic_predictors/rebeca_corr.pdf}
% \caption{Relationship between aesthetic quality and verifier score. Although the correlation is statistically significant, aesthetics explain only a fraction of the variance ($R^2=0.32$), indicating that personalization captures additional signal beyond aesthetic quality.}
% \label{fig:aes-vs-score}
% \end{figure}

% \begin{figure}[h]
% \centering
% \includegraphics[width=0.8\linewidth]{figures/permutation_test/posterior_permtest.pdf}
%     \caption{Permutation test assessing personalization in \texttt{REBECA}. Correctly matched user-image pairs yield verifier scores nearly five standard deviations higher than random assignments, providing strong evidence that \texttt{REBECA} produces genuinely outputs.}
% \label{fig:rebeca-permed}
% \end{figure}

\section{Controllable Preference Navigation}

REBECA embeddings enable controllable preference navigation via spherical interpolation between an original image and a personalized target. Given an image $I_0$ with embedding $z_0=\mathrm{CLIP}(I_0)$ and a REBECA sample $z_1$ for a target user, we generate a trajectory using standard slerp~\cite{showmake_slerp,ramesh2022hierarchicaltextconditionalimagegeneration,levi2025the}:
\[
\mathrm{slerp}(z_0,z_1;t) = \frac{\sin((1-t)\theta)}{\sin\theta}\,\hat{z}_0 + \frac{\sin(t\theta)}{\sin\theta}\,\hat{z}_1,\quad t\in[0,1],
\]
where $\theta$ is the angle between normalized embeddings $\hat{z}_0,\hat{z}_1$. Decoding these interpolated embeddings yields smooth edits that consistently increase the verifier’s predicted preference (\cref{fig:interpolations}), providing a user-aligned control axis orthogonal to text prompts.

%\textbf{Intuition.} Because \texttt{REBECA} learns a generative prior over a user’s preference manifold in embedding space, each point along the interpolation path remains on regions of high personalized likelihood. Moving in the direction of $\mathbf{z}_1$ therefore corresponds to editing the image toward what the user is more likely to prefer, while preserving semantic coherence.

%\textbf{Practical use.} By decoding the interpolated embeddings, we obtain a sequence of images that smoothly transition from the original content to a more personalized variant without explicit textual prompting or fine-tuning (see Figure ~\cref{fig:interpolations}).

%Having statistically verified \texttt{REBECA}'s personalization capabilities, we now \section{\texttt{REBECA} as a personalized image editor}\label{sec:editor}

\section{Discussion}				
\noindent\textbf{Discussion.}
\texttt{REBECA} demonstrates that implicit feedback alone can drive scalable personalization for diffusion models. By learning a lightweight conditional prior over CLIP embeddings and decoding with a frozen backbone, it provides a simple and general framework for connecting recommender signals with generative models—without paired comparisons, per-user fine-tuning, or LLM mediation. Because the method operates in embedding space, it naturally extends to other modalities such as audio, video, or cross-modal generators equipped with pretrained encoders.

Limitations include verifier shift arising from the surrogate preference model, the modest scale of our behavioral dataset, and cold-start scenarios where new users have limited interaction history. Future work includes reinforcement-style preference optimization directly on generative trajectories, large-scale implicit-feedback corpora, and richer multimodal conditioning to extend personalization beyond images.

{
    \small
    \bibliographystyle{ieeenat_fullname}
    \bibliography{main}
}

\appendix
\newpage
\onecolumn
\section{\texttt{REBECA} Prior Architecture}
\label{sup:architecture}
Our conditional diffusion prior is a lightweight and designed for
personalized CLIP embeddings. Given a noisy embedding $I^e_t$ and conditioning, the model predicts the clean embedding $I^e_0$.

We first tokenize the CLIP embedding using a learned tokenizer that maps the
1D embedding into a small set of tokens. Each token is projected to a shared
hidden dimension. Conditioning is injected through three tokens: user, rating, and timestep embeddings. These are mapped
into a conditioning vector via a two-layer MLP.

The core of the model consists of $L$ \emph{PriorBlocks}, each combining:
\begin{itemize}
    \item AdaLN-Zero layers for stable, scale-free conditioning,
    \item self-attention over the image embedding tokens,
    \item  a gated cross-attention mechanism that selectively attends to the
conditioning tokens, and
    \item a zero-initialized MLP residual block for controlled feature updates.

\end{itemize}

All residual pathways are initialized at zero, ensuring training stability and
preventing early over-conditioning.
After $L$ blocks, the model projects tokens back to the original token dimension
and merges them to reconstruct the predicted embedding. The entire architecture contains only 4.4M parameters, so each full training run takes under 10 minutes on a single RTX 4090.
\section{Training Protocol}
\label{sup:training-protocol}
\subsection{Hyperparameter Search}
\textbf{Overview.}
\texttt{REBECA}'s lightweight diffusion prior enables a fully exhaustive hyperparameter search, something typically infeasible for user–conditioned generative models.  
All models were trained on the same hardware (RTX~4090) using identical data splits and random seeds to ensure comparability.

\vspace{0.5em}
\noindent\textbf{Architectures.}
We evaluate a broad family of adapter architectures, including transformer–based variants, cross–attention mechanisms, and a direct residual diffusion prior (\texttt{rdp}), which emerged as the best-performing and most stable model.  
The grid includes variations in depth, attention width, and embedding dimensionality:
\begin{itemize}[nosep,leftmargin=1.2em]
    \item \textbf{Depth:} \{6, 8, 12\} layers  
    \item \textbf{Heads:} \{4, 8\} attention heads  
    \item \textbf{Hidden size:} \{128, 256\}  
    \item \textbf{Token count:} \{16, 32\} (when applicable)
\end{itemize}

\vspace{0.5em}
\noindent\textbf{Diffusion and Objective.}
The prior is trained using several diffusion objectives to test robustness:
\begin{itemize}[nosep,leftmargin=1.2em]
    \item $\epsilon$–prediction  
    \item sample–prediction  
    \item $v$–prediction  
\end{itemize}
We experiment with multiple noise schedules---\texttt{laplace}, \texttt{squaredcos\_cap\_v2}, and \texttt{epsilon}---and fix the number of timesteps to $1000$ to avoid confounding training comparisons.

\vspace{0.5em}
\noindent\textbf{Optimization.}
All models use the \texttt{AdamW} optimizer and a \texttt{ReduceLROnPlateau} scheduler. Hyperparameters were intentionally kept narrow to isolate architectural effects:
\begin{itemize}[nosep,leftmargin=1.2em]
    \item learning rate: $1\times10^{-4}$
    \item batch size: 64  
    \item samples per user: 100  
    \item no gradient clipping or additional normalization
\end{itemize}

\vspace{0.5em}
\noindent\textbf{User Conditioning.}
We sweep over configurations for user thresholding and normalization, though the final model uses no score normalization and no thresholding:
\begin{itemize}[nosep,leftmargin=1.2em]
    \item normalization: \texttt{none}  
    \item user threshold: $0$
\end{itemize}

\vspace{0.5em}
\noindent\textbf{Compute and Stability.}
Every configuration trains in under 25 minutes on a single 4090 GPU, making the full grid search (dozens of runs) computationally tractable.  
This is a major advantage of \texttt{REBECA}'s formulation: the prior is small enough to train repeatedly, allowing principled exploration of design choices rather than relying on heuristics or one-off tuning.

\vspace{0.5em}
\noindent\textbf{Selection Criterion.}
For each run, we evaluate validation loss across objectives. We select the best performing models of each class and inspect a sample of 25 generated images, five for each of the first five users, for image quality.

\subsection{Final Configuration}
\label{sup:final-config}

After completing the exhaustive grid search described above, a single model emerged as the most stable and best-performing across all evaluation metrics: the \texttt{rdp} (\texttt{REBECA} Diffusion Prior) with a lightweight 6-layer architecture.

\paragraph{Backbone.}
All REBECA results use \texttt{Stable-Diffusion~v1.5} as the image generator.  
We load a standard IP-Adapter (\texttt{h94/IP-Adapter}, \texttt{ip-adapter\_sd15.bin}) to provide the visual-conditioning interface, and disable the safety checker for reproducibility.

\paragraph{Winning REBECA Prior.}
The best configuration corresponds to the following hyperparameters:
\begin{itemize}[nosep,leftmargin=1.2em]
    \item \textbf{Architecture:} \texttt{rdp} prior  
    \item \textbf{Layers / Heads / Width:} 6 layers, 8 heads, hidden dimension 128  
    \item \textbf{Tokens:} 32 learned latent tokens  
    \item \textbf{Image embedding dim:} 1024 (CLIP-ViT-bigG)  
    \item \textbf{Users:} 210  
    \item \textbf{Score classes:} 2 (like/dislike)  
\end{itemize}

\paragraph{Diffusion Objective.}
The winning configuration uses:
\begin{itemize}[nosep,leftmargin=1.2em]
    \item \textbf{Prediction type:} \texttt{sample}  
    \item \textbf{Timesteps:} 1000  
    \item \textbf{Noise schedule:} \texttt{squaredcos\_cap\_v2}  
    \item \textbf{Clip-sample:} disabled  
\end{itemize}

The full diffusion scheduler is:
\begin{verbatim}
DDPMScheduler(
    num_train_timesteps=1000,
    beta_schedule="squaredcos_cap_v2",
    clip_sample=False,
    prediction_type="sample"
)
\end{verbatim}

\paragraph{Weights.}
All final experiments load the trained prior from:
\begin{verbatim}
comprehensive_study_20250830_013540/
    modelrdp_num_layers6_num_heads8_hidden_dim128_tokens32_
    lr0.0001_optadamw_schreduce_on_plateau_bs64_
    nssquaredcos_cap_v2_ts1000_spu100_csFalse_objsample_normnone_uthr0
\end{verbatim}

\paragraph{Compute.}
This model trains in under $\sim$10 minutes on a single RTX~4090, which enables the exhaustive search strategy and provides a key practical advantage over existing personalization pipelines.

\section{Image Tagging Pipeline}
\label{sup:image-tagging}
We employ the open-source checkpoint \texttt{llava-hf/llava-1.5-7b-hf} loaded with \texttt{transformers}.
Each image is processed with the following fixed instructions:

\begin{quote}
\begin{verbatim}
You are a tagging engine. Return ONLY a JSON object with fields:
{"caption": str, "objects": [str], "attributes": [str],
 "styles": [str], "colors": [str]}.
Rules: lowercase; <=10 items total; no nulls.
\end{verbatim}
\end{quote}

\section{Baseline Specification}
\subsection{LoRA per User}
\label{suppl:lpu}
We fit a single adapter per user and we employ each user's liked images as training data. Due to the variability in dataset sizes, we must adapt the training configuration for each user depending on the number of liked images. See \cref{tab:lora_hyperparams}
for the various configurations. We adapt Hugging Face's Parameter Efficient Fine-Tuning (\textit{PeFT}) \cite{peft} script for Stable Diffusion \cite{rombach2021highresolution} to filter by user tags and loop over their IDs and configurations for training. 

\begin{table}
\centering
\small
\setlength{\tabcolsep}{6pt}
\renewcommand{\arraystretch}{1.2}
\caption{\textbf{Per-user LoRA training hyperparameters.} The LoRA rank ($r$), scaling factor ($\alpha$), dropout, training steps, and warmup schedule are adapted based on the number of available user images ($N_i$).}
\label{tab:lora_hyperparams}
\begin{tabular}{c c c c c c}
\toprule
\textbf{$N_i$ range} & \textbf{LoRA $r$} & \textbf{LoRA $\alpha$} & \textbf{Dropout} & \textbf{Steps} & \textbf{Warmup} \\
\midrule
$N_i < 8$           & 8  & 8  & 0.10 & 1200 & 100  \\
$8 \le N_i \le 24$  & 16 & 16 & 0.07 & 1500 & 120  \\
$25 \le N_i \le 60$ & 32 & 32 & 0.05 & 1700 & 150  \\
$61 \le N_i \le 120$& 64 & 64 & 0.05 & 2000 & 200  \\
$N_i > 120$         & 128 & 128 & 0.05 & 2500 & 200  \\
\bottomrule
\end{tabular}
\end{table}

\subsection{Shared LoRA}
\label{suppl:shared-lora}
Collaborative filtering is a fundamental concept in Recommender Systems, and it posits a common latent representation for users. In this spirit, a single LoRA model with rank $r=512$ is calibrated with all users simultaneously. For training hyperparameters, see Table \ref{tab:shared_lora_hyperparams}. 

\begin{table}
\centering
\small
\setlength{\tabcolsep}{6pt}
\renewcommand{\arraystretch}{1.2}
\caption{\textbf{Shared LoRA training hyperparameters.} The shared LoRA rank ($r$), scaling factor ($\alpha$), dropout, training steps, and warmup schedule.}
\label{tab:shared_lora_hyperparams}
\begin{tabular}{ c c c c c}
\toprule
 \textbf{LoRA $r$} & \textbf{LoRA $\alpha$} & \textbf{Dropout} & \textbf{Steps} & \textbf{Warmup} \\
\midrule
  512  & 512  & 0.05 & 10000 & 1000  \\
\bottomrule
\end{tabular}
\end{table}

\subsection{VLM-Persona Generation}
\label{sup:vlm-persona}
\paragraph{Prompt Construction.}
Each user's persona defines positive and negative keyword lists.
When few positive terms exist, fallback categories (``nature, landscape, portrait, cityscape, animals'') are used.  
The builder cycles deterministically through available terms and appends either
the user's persona or a default stylistic tail.  

\paragraph{Implementation.}
We employ the \texttt{Diffusers} library for inference in half precision (fp16) on GPU, 
with the safety checker disabled for consistent reproducibility.

\vspace{0.4em}
\begin{table}[h]
\centering
\small
\setlength{\tabcolsep}{5pt}
\renewcommand{\arraystretch}{1.1}
\caption{\textbf{Fallback prompt components} used when user profiles lack sufficient detail.}
\label{tab:persona_fallbacks}
\begin{tabular}{l p{0.65\linewidth}}
\toprule
\textbf{Type} & \textbf{Default values or description} \\
\midrule
Positive keywords & \texttt{\{nature, landscape, portrait, cityscape, animals\}} \\
Style suffix & ``high quality, detailed, natural lighting'' \\
Negative prompt & \texttt{\{low quality, blurry, deformed, overexposed, underexposed\}} \\
\bottomrule
\end{tabular}
\end{table}

\vspace{-0.5em}
\begin{algorithm}[h]
\caption{Prompt Builder for User $u_i$}
\label{alg:prompt_builder}
\begin{algorithmic}[1]
\State \textbf{Input:} profile $(\text{persona}, \text{pos}, \text{neg})$, image index $j$
\State $k \gets$ number of positive terms to include (1–2)
\State $P \gets$ cycle through $\text{pos}$ deterministically for $k$ terms
\If{$P$ is empty} \State $P \gets$ random fallback from Table~\ref{tab:persona_fallbacks}
\EndIf
\State $\text{style} \gets$ persona text if available else default suffix
\State $\text{prompt} \gets P + \text{style}$
\State $\text{negprompt} \gets$ user negatives or default negatives
\State \Return $(\text{prompt}, \text{negprompt})$
\end{algorithmic}
\end{algorithm}

\paragraph{Generation Summary.}
\begin{itemize}[leftmargin=1.5em]
    \item \textbf{Model:} Stable Diffusion v1.5 (\texttt{stable-diffusion-v1-5})
    \item \textbf{Inference steps:} 50
    \item \textbf{Guidance scale:} 5.0
    \item \textbf{Images per user:} 25
    \item \textbf{Total users:} 210
    \item \textbf{Seeding:} $\text{seed}(u_i, j) = 42 + 10{,}000\times i + j$
    \item \textbf{Output:} Serialized image bundles per user (\texttt{.imgs})
\end{itemize}

\vspace{0.5em}
This procedure, implemented in \texttt{vlm\_personas.ipynb}, serves as the
\textit{text-conditioned personalization baseline} for comparison against
LoRA-based and diffusion-prior models.

\section{Verifier diagnostics}\label{sec:diag}
\begin{figure}[h]
\centering
\includegraphics[width=\linewidth]{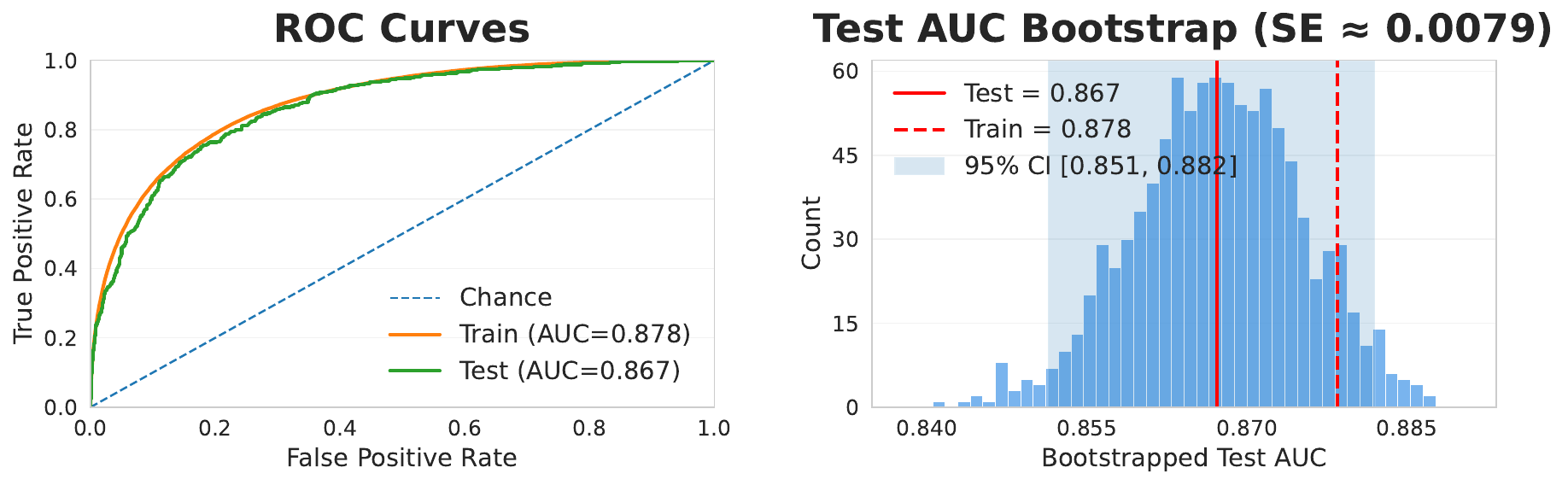}
\caption{Verifier performance on the training and test sets. The two curves are nearly indistinguishable, and the ROC-AUC values are not statistically different.}
\label{fig:diag}
\end{figure}

Figure~\ref{fig:diag} shows that the verifier achieves nearly identical performance on the training and test sets. In the second panel, we report the results of a bootstrap analysis of the test data, demonstrating that the training ROC-AUC lies within two standard errors of the test ROC-AUC. This indicates that the verifier does not overfit and generalizes well to unseen samples.

\section{Formally testing for personalization}\label{sec:hyp-test}

Algorithm \ref{alg:perf-eval} implements a permutation test used to formally verify the personalization aspect of \texttt{REBECA}. In Algorithm \ref{alg:perf-eval}, $\widehat{\text{Score}}$ is the median verifier score across users and $\widehat{\text{Score}_b}$ is the median verifier score across users after permutation using the random seed equals $b$.
\begin{algorithm}[H]
   \caption{Testing for \texttt{REBECA} personalization}
   \label{alg:perf-eval}
\begin{algorithmic}[1]
   \Require Users, \texttt{REBECA} model, verifier \(\hat{v}\), significance level \(\alpha\), number of permutations \(B=1000\)
   \For{each user}
       \State Generate 30 images using \texttt{REBECA}
   \EndFor
   \State Compute baseline performance \(\widehat{\text{Score}}\) using verifier \(\hat{v}\)
   \For{\(b = 1\) to \(B\)}
       \State Randomly permute generated images across users
       \State Compute \(\widehat{\text{Score}}_b\) using verifier \(\hat{v}\)
   \EndFor
   \State Compute p-value: 
   \[
       p = \frac{1 + \sum_{b=1}^B \mathds{1}\left[\widehat{\text{Score}} \le \widehat{\text{Score}}_b\right]}{B+1}
   \]
   \If{\(p \le \alpha\)}
       \State Reject null hypothesis \(H_0: U \ind I\) 
       \State (that \texttt{REBECA}'s images do not depend on users)
   \EndIf
\end{algorithmic}
\end{algorithm}

\section{Extra plots for the personalization results}\label{sec:personalization}

\begin{figure}[t]
\centering
\includegraphics[width=0.95\linewidth]{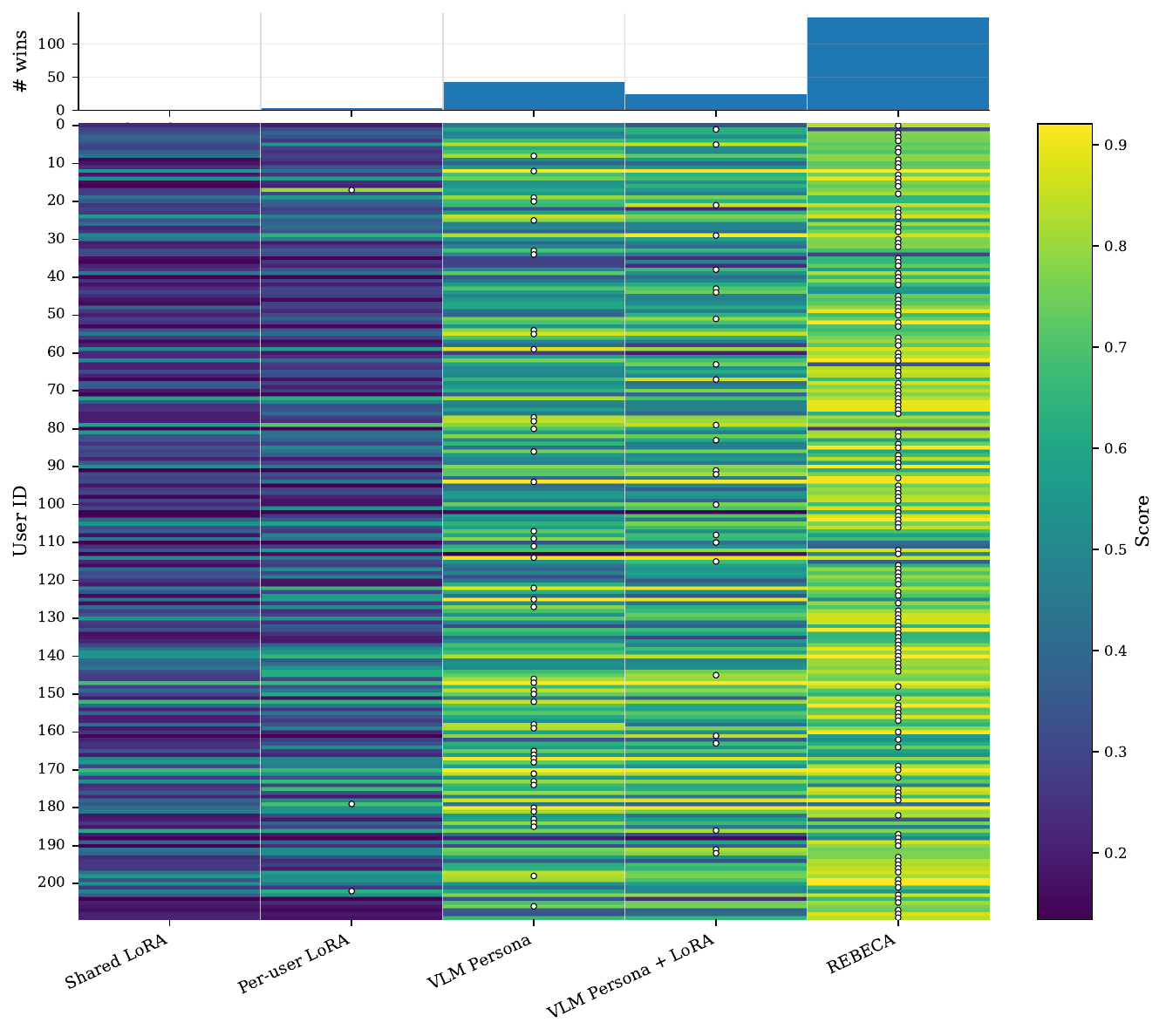}
\caption{Verifier scores for each user across different generation methods. \texttt{REBECA} achieves the highest scores for most users, indicating stronger personalization performance.}
\label{fig:disag-score}
\end{figure}

\begin{figure}[t]
\centering
\includegraphics[width=0.95\linewidth]{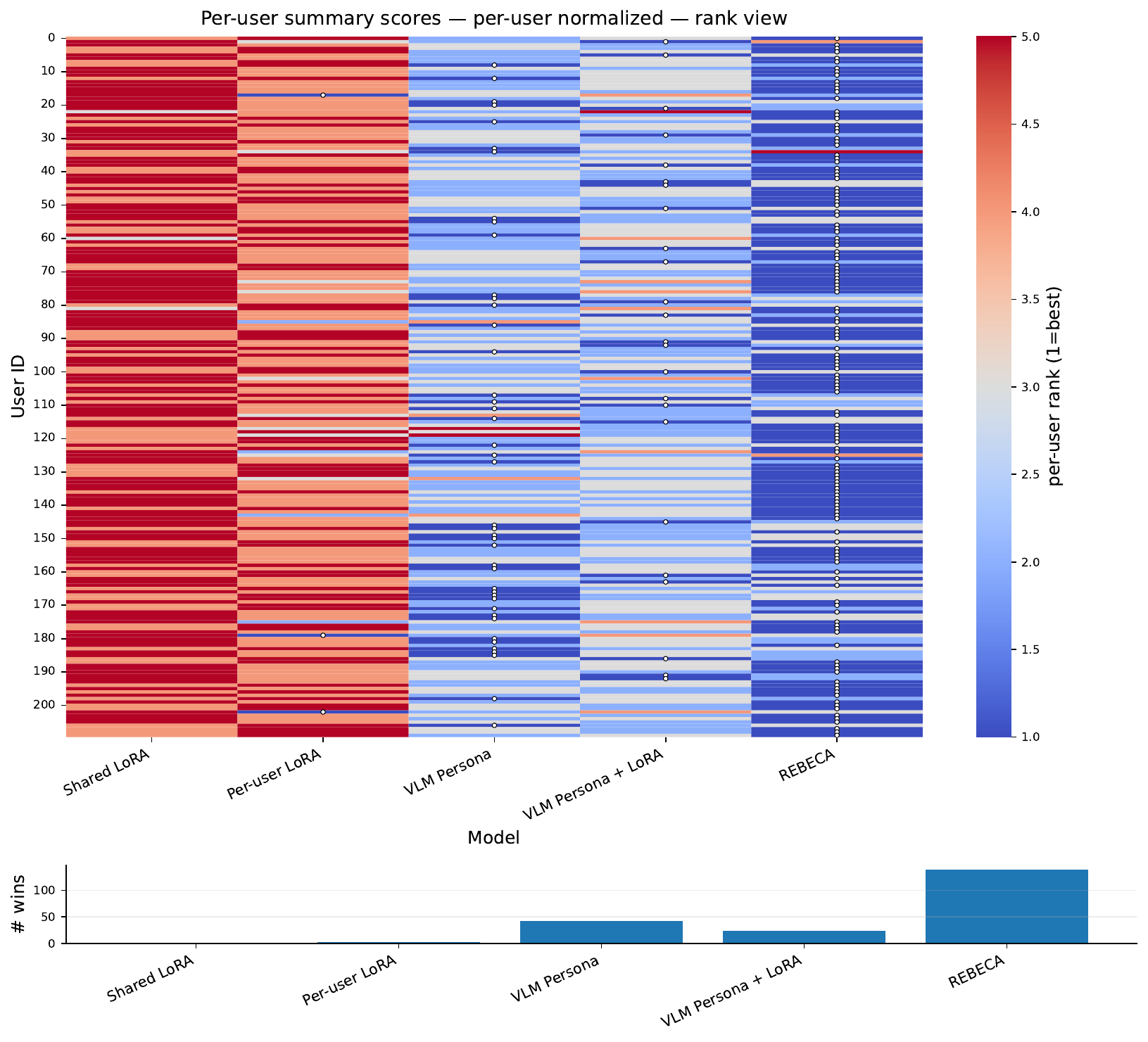}
\caption{Relative ranking of generation methods per user. \texttt{REBECA} ranks highest for the majority of users, outperforming all baselines in personalized generation quality.}
\label{fig:disag-rank}
\end{figure}

Figures \ref{fig:disag-score} and \ref{fig:disag-rank} present the performance of each method across users in terms of verifier scores and relative rankings, respectively. In both cases, \texttt{REBECA} consistently outperforms all baseline methods for the vast majority of users.

\section{Extra plots for the aesthetics and overall quality results}\label{sec:aes-quality}
\begin{figure}[t]
\centering
\includegraphics[width=0.9\linewidth]{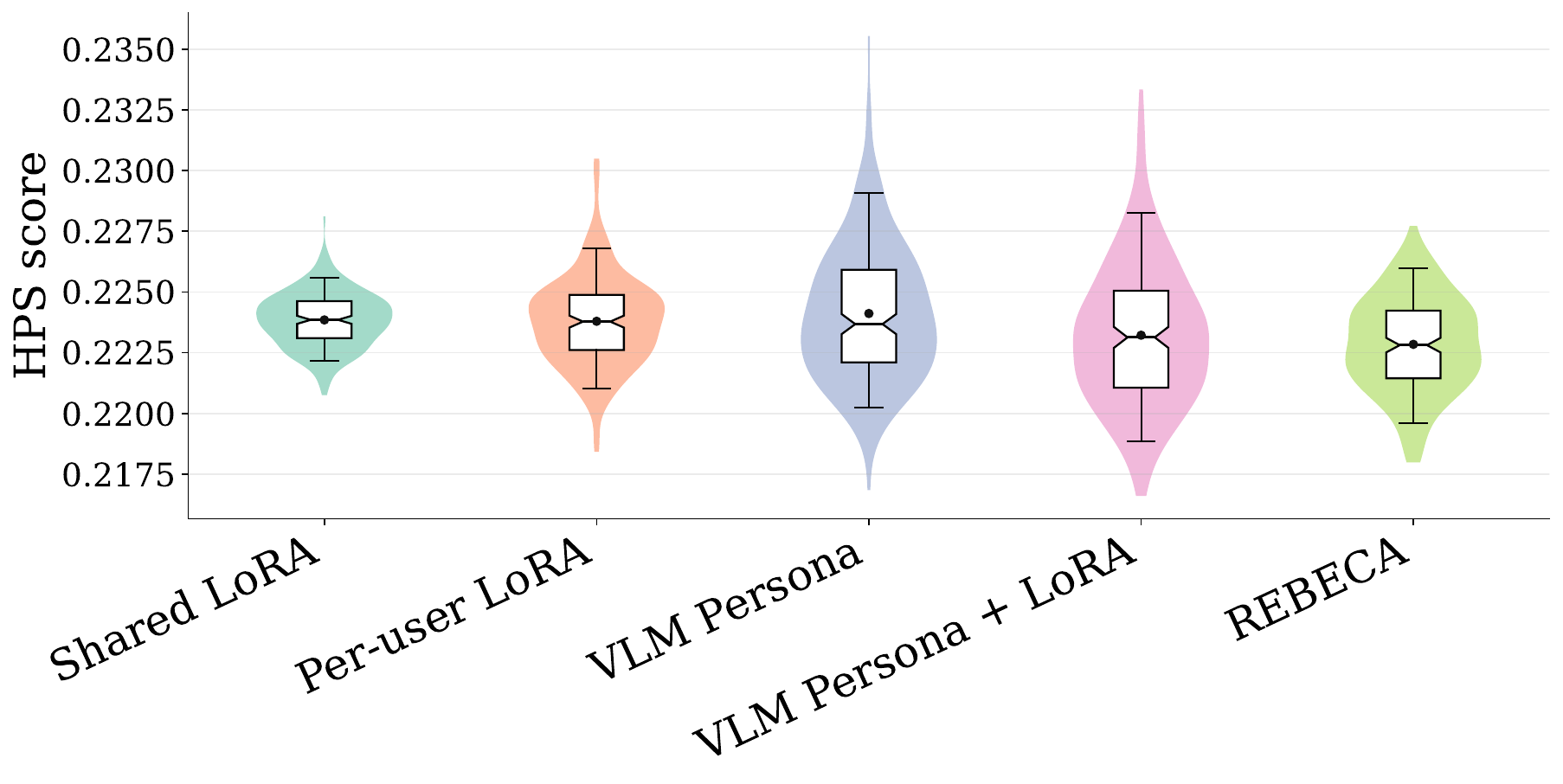}
\caption{\textbf{HPSv2 results under the zero-prompt setting.}
When no textual prompt is provided, all models achieve similar overall quality scores, indicating that prompt content mainly influences alignment rather than visual fidelity. \texttt{REBECA} maintains competitive quality while preserving personalization robustness.}
\label{fig:zero-prompt-hps}
\end{figure}

Figure \ref{fig:zero-prompt-hps} reports the HPSv2 results obtained when using an empty prompt. In this zero-prompt setting, all generative methods display comparable overall quality. Notably, \texttt{REBECA} is capable of delivering personalized generations without hurting overall generation quality.

\section{Ablations}
\subsection{System Prompts}
\label{sup:system-prompts}
To support the prompt–control ablation reported in the main paper, we evaluate \texttt{REBECA} under three system–prompt levels \(c_t\), ranging from no prompt to strongly opinionated aesthetic prompts with aggressive negative prompts. For all experiments, we freeze the best-performing diffusion-prior checkpoint and generate personalized embeddings for all 210 users, which are then decoded using a Stable Diffusion v1.5 pipeline with an IP-Adapter (identical sampling settings across conditions).

The three prompt levels used in the ablation are:

\paragraph{Level 0: No prompt.}
\begin{itemize}
    \item \textbf{Positive prompt:} \texttt{""}
    \item \textbf{Negative prompt:} \texttt{""}
\end{itemize}

\paragraph{Level 1: Mild quality prompt.}
\begin{itemize}
    \item \textbf{Positive prompt:}
    \texttt{"high quality photo"}
    \item \textbf{Negative prompt:}
    \texttt{"bad quality photo, letters"}
\end{itemize}

\paragraph{Level 2: Strong descriptive/negative prompts.}
\begin{itemize}
    \item \textbf{Positive prompt:}
    \texttt{"Realistic image, finely detailed, with balanced composition and harmonious elements. Dynamic yet subtle tones, versatile style adaptable to diverse themes and aesthetics, prioritizing clarity and authenticity."}
    \item \textbf{Negative prompt:}
    \texttt{"deformed, ugly, wrong proportion, frame, watermark, low res, bad anatomy, worst quality, low quality"}
\end{itemize}

For each prompt level, we generate 10 personalized images per user (2{,}100 images per condition). The outputs are subsequently evaluated by our verifier model. As discussed in the main text, we observe \emph{no statistically significant effect} of prompt level \(c_t\) at any \texttt{REBECA} CFG value (Fig.~\ref{fig:ablations-cross-prompts}), indicating that personalization arises from the diffusion prior rather than from prompt engineering.

% Commented out due t arxiv
% \section{Samples}

% In this Section, we add samples for the first ten users by ID in our dataset. We generate the images below by fixing the seed and increasing the prior CFG.

% \begin{figure}[t]
% \centering
% \includegraphics[width=\linewidth]{figures/samples/rebeca_users12_cols10_embcfg3.0.pdf}
% \caption{Sample images for users 1--12 with prior CFG 3.0.}
% \label{fig:samples-cfg-3}
% \end{figure}

% \begin{figure}[t]
% \centering
% \includegraphics[width=\linewidth]{figures/samples/rebeca_users12_cols10_embcfg5.0.pdf}
% \caption{Sample images for users 1--12 with prior CFG 5.0.}
% \label{fig:samples-cfg-5}
% \end{figure}

% \begin{figure}[t]
% \centering
% \includegraphics[width=\linewidth]{figures/samples/rebeca_users12_cols10_embcfg7.0.pdf}
% \caption{Sample images for users 1--12 with prior CFG 7.0.}
% \label{fig:samples-cfg-7}
% \end{figure}

% \begin{figure}[t]
% \centering
% \includegraphics[width=\linewidth]{figures/samples/rebeca_users12_cols10_embcfg9.0.pdf}
% \caption{Sample images for users 1--12 with prior CFG 9.0.}
% \label{fig:samples-cfg-9}
% \end{figure}

% With a lower CFG, the images are more diverse, capturing broader regions of the user preference manifold (Figure \ref{fig:samples-cfg-3}). Hence, the higher recall results in \cref{tab:pr-at-k} at lower CFG values. As CFG increases, sampling concentrates around fewer high preference regions, and generations become more demarcated across users (Figure \ref{fig:samples-cfg-9}). 

\end{document}